\documentclass[twocolumn,showpacs,preprintnumbers,amsmath,amssymb,epsfig,widetext]{revtex4}

\usepackage{graphicx}
\usepackage{dcolumn}
\usepackage{bm}
\usepackage{epsfig}
\newcommand{\la}{\lesssim}
\newcommand{\ga}{\gtrsim}

\input epsf

\def\be{\begin{equation}}
\def\ee{\end{equation}}
\def\ba{\begin{eqnarray}}
\def\ea{\end{eqnarray}}

\newcommand{\mA}{{\cal A}}
\newcommand{\mB}{{\cal B}}
\newcommand{\ds}{\mathrm{d}}

\begin{document}

\preprint{}

\title{Models of $f(R)$ Cosmic Acceleration that Evade Solar-System Tests}

\author{Wayne Hu$^{1,2}$ and Ignacy Sawicki$^{1,3}$}
\email{whu@background.uchicago.edu}
\affiliation{{}$^1$ Kavli Institute for Cosmological Physics, Enrico Fermi
Institute,  University of Chicago, Chicago IL 60637 \\
{}$^2$ Department of Astronomy \& Astrophysics,  University of Chicago, Chicago IL 60637\\
{}$^3$ Department of Physics,  University of Chicago, Chicago IL 60637
}

\date{\today}

\begin{abstract}
We study a class of metric-variation
$f(R)$ models that accelerates the expansion without a cosmological
constant and satisfies both cosmological and solar-system tests in the small-field limit
of the parameter space.  Solar-system tests {\it alone} place only weak bounds on these
models, since the additional scalar degree of freedom is locked to the high-curvature
general-relativistic prediction across more than 25 orders of magnitude in density,
out through the solar corona.  This agreement requires that the galactic
halo be of sufficient extent to maintain the galaxy at high curvature in the presence
of the low-curvature cosmological background.   If the galactic halo and local environment
in $f(R)$ models
do not have substantially deeper potentials than expected in $\Lambda$CDM, then
cosmological field amplitudes $|f_{R}| \ga 10^{-6}$ will cause the galactic interior
to evolve to low curvature during the acceleration epoch.
Viability of large-deviation models therefore rests on the structure and evolution of
the galactic halo, requiring cosmological simulations of $f(R)$ models,
and not directly on solar-system tests.   Even small deviations that conservatively satisfy
both galactic and solar-system constraints can still be tested by future,
percent-level measurements
of the linear power spectrum, while they remain undetectable to cosmological-distance measures.
Although we illustrate these effects in a specific class of  models,
the requirements on $f(R)$ are phrased in a nearly model-independent manner.
\end{abstract}



\maketitle

\section{Introduction}
\label{sec:introduction}

Cosmic acceleration, in principle, can arise not from dark energy---a new, exotic form of matter---but rather from a modification of gravity that appears on large scales.
The addition of a non-linear function of the Ricci scalar $R$
to the Einstein-Hilbert action has been demonstrated to cause acceleration for a wide variety of $f(R)$ functions \cite{Caretal03,Nojiri:2006ri,Capozziello:2002rd,Capozziello:2003tk,
Nojiri:2003ft,Nojiri:2003rz,Faraoni:2005vk,delaCruz-Dombriz:2006fj,Poplawski:2006kv,
Brookfield:2006mq,Li:2006ag,Sotiriou:2006qn,Sotiriou:2006sf,
Sotiriou:2006hs,Bean:2006up,Baghram:2007df,Bazeia:2007jj,Li:2007xn,Bludman:2007kg,
Rador:2007wq,Rador:2007gq,Sokolowski:2007pk,Faraoni:2007yn,
Faraoni:2006sy,Nojiri:2003ni,Wang:2004vs,Meng:2003sx,Abdalla:2004sw,
Cognola:2005de,Capozziello:2005ku,Allemandi:2005qs,Koivisto:2005yc,Clifton:2005aj,
Mena:2005ta,Amarzguioui:2005zq,Brevik:2006md,Koivisto:2006ie,PerezBergliaffa:2006ni,
Cognola:2007vq,Capozziello:2007gm,Nojiri:2006gh,Nojiri:2006su,Capozziello:2006dj,
Fay:2007uy,Fay:2007gg,Nojiri:2005vv,Sami:2005zc,Calcagni:2005im,
Tsujikawa:2006ph,Guo:2006ct,Sanyal:2006wi,Leith:2007bu,Carter:2005fu,
Koivisto:2006ai,Koivisto:2006xf,Nojiri:2006je,Nojiri:2006be,Cognola:2006sp,
Nojiri:2005jg,Cognola:2006eg,Nojiri:2007te,SonHuSaw06}.

What is less clear in the literature is whether any proposed metric-variation
 $f(R)$ modification can simultaneously satisfy stringent solar-system bounds on deviations from general relativity as well as accelerate the expansion at late times \cite{MulVil06a,Woo06,Sot05,Cem05,Allemandi:2005tg,
Ruggiero:2006qv,Sotiriou:2006pq,
ShaCaiWanSu06,Bustelo:2006ms,Olmo:2006eh,Zha07,Kainulainen:2007bt}.  Chiba \cite{Chi03} showed that the fundamental difficulty is that $f(R)$ gravity introduces a scalar degree of freedom with the same coupling to matter as gravity that, at the background cosmological density, is extremely light.  This light degree of freedom produces a
  long-range fifth force or, equivalently, a dissociation of the curvature of the space-time from the local density. As a result, the metric around the Sun is predicted to be different than is implied by observations.  This problem has been explicitly proven to exist for a wide variety of $f(R)$ models, if the Sun is placed into a background of cosmological density \cite{EriSmiKam06,Chiba:2006jp}.

If high density could be re-associated with high curvature this difficulty would disappear.  The scalar degree of freedom would become massive in the high-density solar vicinity and hidden from solar-system tests by the so-called chameleon mechanism \cite{KhoWel04,NavAco06,FauTegBunMao06}.  This requires a form for $f(R)$ where
the mass squared of the scalar is large and positive at high curvature \cite{SonHuSaw06}.  Such a condition is also required for agreement with high-redshift cosmological tests from
the cosmic microwave background (CMB) \cite{AmePolTsu06a,SawHu07}.   It should be considered as a necessary condition for a successful $f(R)$ model; it is violated in the original inverse-curvature model and many other generalizations (e.g. \cite{Zha05}).

Faulkner {\it et al.}\ \cite{FauTegBunMao06} analyzed a class of models where solar-system tests of gravity could be evaded, but only at the price of reintroducing the cosmological constant as a constant piece of
$f(R)$ that drives the cosmic acceleration but is unrelated to local modifications of gravity. Moreover, for these models to satisfy local constraints, all aspects of the cosmology are essentially
indistinguishable from general relativity with a cosmological constant.

In this {\it Paper}, we introduce a class of $f(R)$ models that do not contain a cosmological constant and yet are explicitly designed to satisfy cosmological and solar-system constraints in certain limits of parameter space.  We use these models to ask under what circumstances it is possible to significantly modify cosmological predictions and yet evade all local tests of gravity.

We begin in \S \ref{sec:cosmology} by introducing the model class, its effect on the background expansion history and the growth of structure.   We show that cosmological tests of the growth of structure can, in principle, provide extremely precise tests of $f(R)$ gravity that rival local constraints and complement them in a very different range in curvature. We then analyze local tests of gravity in \S\ref{sec:local} and show that solar-system tests {\it alone} are fairly easy to evade, provided gravity behaves similarly to general relativity in the galaxy.  However, if cosmological deviations from general relativity are required to be large, the latter condition is satisfied only with extreme and testable changes to the galactic halo. We discuss these results in \S \ref{sec:discussion}.

\section{$f(R)$ Cosmology}
\label{sec:cosmology}

In this section, we discuss the cosmological impact of $f(R)$ models of the acceleration.  We begin in \S \ref{sec:model} by introducing a class of models that
 accelerate the expansion without a true cosmological constant but nonetheless includes the phenomenology of $\Lambda$CDM as a limiting case.   We then
describe the background equations of motion (\S \ref{sec:background}) and their representation as an equation for the scalar degree of freedom (\S \ref{sec:field}). Finally, we calculate the expansion history (\S \ref{sec:expansion})
and linear power spectrum (\S \ref{sec:linear}) in our class of $f(R)$ models.

\subsection{Model}
\label{sec:model}

We consider a modification to the Einstein-Hilbert action of the form \cite{Sta80}
\begin{equation}
S=\int \ds^4x \sqrt{-g}\left[ {R+f(R) \over 2 \kappa^2}+{\cal L}_{\rm m} \right]\,,
\label{eqn:action}
\end{equation}
where $R$ is the Ricci scalar,  which we will refer to as
the curvature, $\kappa^{2}\equiv 8\pi G$, and ${\cal L}_m$ is the matter Lagrangian. Note that a constant $f$ is simply a cosmological constant.  We work in the Jordan frame throughout this paper.

We choose the functional form of $f(R)$ to satisfy certain observationally desirable properties. Firstly, the cosmology should mimic $\Lambda$CDM in the high-redshift regime where it is well-tested by the CMB.  Secondly, it should accelerate the expansion at low redshift with an expansion history that is close to $\Lambda$CDM, but without a true cosmological constant. Thirdly, there should be sufficient degrees of freedom in the parametrization to encompass as broad a range of low-redshift phenomena as is currently observationally acceptable. Finally, for the purposes of constraining small deviations from general relativity with cosmological and solar-system tests, it should include the phenomenology of $\Lambda$CDM as a limiting case.

These requirements suggest that we take
\begin{eqnarray}
\lim_{R\rightarrow \infty} f(R) &=& {\rm const.} \,,\nonumber\\
\lim_{R\rightarrow 0} f(R) &=& 0\,,
\end{eqnarray}
which can be satisfied by a general class of broken power law models
\begin{eqnarray}
f(R)& =&- m^2 {   c_1  (R/m^2)^n \over c_2 (R/m^2)^n +1}\,,
\label{eqn:model}
\end{eqnarray}
with $n>0$, and for convenience we take the mass scale
\begin{equation}
m^{2} \equiv {\kappa^{2}\bar\rho_0\over 3} = (8315 {\rm Mpc})^{-2} \left( {\Omega_m h^2 \over 0.13}\right)\,,
\end{equation}
where $\bar\rho_0 = \bar \rho(\ln a=0)$ is the average density today.
$c_1$ and $c_2$ are dimensionless parameters.    It is useful to note that
\begin{equation}
{\kappa^2 \rho \over m^2} = 1.228 \times 10^{30} \left( {\rho \over 1 {\rm g\ cm}^{-3}} \right)
\left( {\Omega_{m}h^{2} \over 0.13} \right)^{-1}\,.
\end{equation}

\begin{figure}[tb]\begin{centering}
\includegraphics[width=0.9\columnwidth]{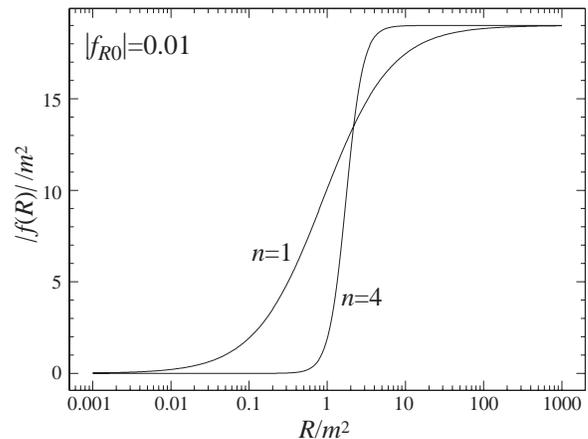}
\caption{Functional form of $f(R)$ for $n=1,4$, with normalization parameters $c_1,c_2$
given by
 $|f_{R0}|=0.01$ and a matching to $\Lambda$CDM densities (see \S \ref{sec:expansion}).
 These functions transition from zero to a constant as $R$ exceeds $m^{2}$.
 The sharpness of the transition increases with $n$ and
 its position increases with $|f_{R0}|$.   During cosmological expansion, the background only reaches
 $R/m^2 \sim 40$ for $|f_{R0}| \ll 1$ and so the functional dependence for smaller
 $R/m^2$ has no impact on the phenomenology.}
\label{fig:model}
\end{centering}\end{figure}

The sign of $f(R)$ is chosen so that its second derivative
\begin{equation}
f_{RR} \equiv {\ds^2 f(R) \over \ds R^2} >0
\end{equation}
for $R \gg m^2$, to ensure that, at high density, the solution is stable at high-curvature \cite{SonHuSaw06}. This condition also implies that cosmological tests at high redshift remain the same as
in general relativity (GR).   For example, the physical matter density $\Omega_m h^2$ inferred from the CMB using GR remains valid for the $f(R)$ models. As such, $m$ is a better choice of scale than
$H_0$ since it does not vary for $f(R)$ models in this class.
A few examples of the $f(R)$ functions are
shown in Fig.~\ref{fig:model}.

There is no true cosmological constant introduced in this class, unlike in the models of \cite{FauTegBun06}.  However, at curvatures high compared with $m^2$, $f(R)$ may be expanded as
\begin{equation}
\lim_{m^{2}/R \rightarrow 0} f(R) \approx -{c_{1}\over c_{2}} m^{2} + {c_{1}\over c_{2}^{2}}
m^{2} \left({m^{2}\over R} \right)^{n} \,.
\label{eqn:taylor}
\end{equation}
Thus the limiting case of $c_1/c_2^2 \rightarrow 0$ at fixed $c_1/c_2$ is a cosmological constant in both cosmological and local tests of gravity, as we shall see.  Moreover, at finite $c_1/c_2^2$, the curvature freezes into a fixed value and ceases to decline with the matter density, creating a class of models that accelerate in a manner similar to $\Lambda$CDM.
These models therefore also do not exhibit the problems of ``mCDTT`'' models with the form
$f(R) = \mu^{4}/R$.  (Note the sign difference from the original inverse curvature CDTT
model \cite{Caretal03}.)  While these models can accelerate the expansion, they evolve in the
future
into an unstable regime where $1+f_{R} < 0$ and also do not contain $\Lambda$CDM as
a limiting case of the parameter space \cite{SawHu07}.
   Note that $n=1$ resembles the mCDTT plus a cosmological
constant at high curvature.  Likewise $n=2$ resembles the inverse curvature
squared model \cite{Mena:2005ta} plus a cosmological constant.

\subsection{Background Evolution Equations}
\label{sec:background}

Variation of the action (\ref{eqn:action}) with respect to the metric yields the modified Einstein equations
\begin{eqnarray}\label{eqn:metricvar}
G_{\alpha\beta} +
f_{R} R_{\alpha\beta}-\left({f\over2} -\Box f_{R}\right) g_{\alpha\beta}
- \nabla_{\alpha}\nabla_{\beta}f_{R}
= \kappa^{2} T_{\alpha\beta}\,,
\end{eqnarray}
where the field,
\begin{equation}
f_R \equiv {\ds f(R) \over \ds R} \,,
\end{equation}
will play a central role in the analyses below.

Since modifications only appear at low redshift, we take a matter-dominated stress-energy tensor. For the background Friedmann-Robertson-Walker (FRW) metric,
\begin{equation}
R =12 H^{2} + 6 H H' \,,
\label{eqn:ricci}
\end{equation}
where $H(\ln a)$ is the Hubble parameter and
$' \equiv \ds/\ds\ln a$.  The modified Einstein equations become the
modified Friedmann equation
\begin{equation}
H^2 - f_R (H H' + H^2) + {1\over 6} f + H^2 f_{RR} R' = {\kappa^2\bar \rho \over 3} \,.
\label{eqn:friedmann}
\end{equation}

To solve these equations, we re-express them in terms of parameters
the values of which vanish in the high-redshift limit where $f(R)$ modifications are negligible
 \begin{eqnarray}
 y_{H} &\equiv& {H^{2}\over m^{2}} - a^{-3} \,,\nonumber\\
y_{R} &\equiv&  {R \over m^{2}} - 3 a^{-3} \,.
\end{eqnarray}
Equations  (\ref{eqn:ricci}) and (\ref{eqn:friedmann}) become a coupled set of
ordinary differential equations
\begin{eqnarray}
y_{H}' &=& {1\over 3} y_{R} - 4 y_{H} \,, \label{eqn:reducedfriedmanna}\\
y_{R}' &=& 9 a^{-3} -{1 \over y_{H}+a^{-3}} {1 \over m^{2}f_{RR}}
\label{eqn:reducedfriedmannb}\\
&& \times
\left[ y_{H} - f_{R}\left( {1 \over 6} y_{R} - y_{H} - {1\over 2}a^{-3}\right) +
{1 \over 6}{f \over m^{2}} \right] \,. \nonumber
\end{eqnarray}
To complete this system, we take the initial conditions at high redshift to be given by detailed balance
of perturbative corrections to $R =\kappa^2\rho$.

The impact of $f(R)$ on the expansion history can be recast as an effective equation of
state for a dark energy model with the same history
\begin{equation}
1+w_{\rm eff} =-{1 \over 3 } {y_H' \over y_H} \,.
\label{eqn:weff}
\end{equation}

The two equations  (\ref{eqn:reducedfriedmanna}) and (\ref{eqn:reducedfriedmannb}) combine  in the high curvature limit---$a^{-3} \gg y_{H}$,$y_{R}$---to form
\begin{equation}
y_{H}'' + [ \ldots ]y_{H}' +{1\over 3 m^{2}f_{RR} a^{-3}} y_{H} = {\rm driving\ terms}
\label{eqn:oscillator} \,,
\end{equation}
where $[\ldots]$ contains a time-dependent friction term whose exact nature is
not relevant for the qualitative argument and the driving terms involve the matter density.
As shown in \cite{SonHuSaw06}, a critical requirement of the model is that
$f_{RR} > 0$ such that Eqn.~(\ref{eqn:oscillator}) becomes an oscillator equation
with real---not imaginary---mass.
 For the background, it is convenient to express this as the
dimensionless quantity
\begin{eqnarray}
B&=& {f_{RR} \over 1+f_R} {R'}{H \over H'} \,.
\end{eqnarray}
This oscillator equation and the parameter $B$ have a simple interpretation in terms of
the scalar field $f_R$ as we shall now see.

 \subsection{Field Equations}
 \label{sec:field}

The impact of $f(R)$ can be alternatively viewed in terms of the field equation for
$f_R$.
The
 trace of Eqn.~(\ref{eqn:metricvar}) can be interpreted as the equation of motion for
 $f_R$
\begin{equation}
 3\Box f_R - R + f_R R - 2 f = -\kappa^2 \rho \,.
 \label{eqn:trace}
\end{equation}
This equation can be recast in the form
\begin{equation}
\Box f_R =  {\partial V_{\rm eff} \over \partial f_R}\,,
\end{equation}
with the effective potential
\begin{equation}
{\partial V_{\rm eff} \over \partial f_R} = {1\over 3}\left( R- f_R R + 2 f - \kappa^2\rho \right)\,. \label{eqn:potn}
\end{equation}

The effective potential has an extremum at
 \begin{equation}
 R - R f_R + 2 f = \kappa^2 \rho \,.
 \end{equation}
In the high-curvature regime, where $|f_R| \ll1$ and $|f/R|\ll 1$, the extremum lies at the general-relativistic  expectation of $R = \kappa^2 \rho$.
The curvature at the extremum is given by
 \begin{equation}
m^2_{f_R}= {\partial^2 V_{\rm eff} \over \partial f_R^2} = { 1\over 3} \left( {{1 + f_R \over f_{RR}} - R} \right)
 \end{equation}
 and hence the extremum is a minimum for $B>0$ and a maximum for $B<0$ in
 the high-curvature limit with $|f_R|, |f_{RR} R| \ll 1$.
   Finally the Compton wavelength
 \begin{equation}
 \lambda_{f_R} \equiv m_{f_R}^{-1} \,,
 \end{equation}
 implies that, in this limit,
 \begin{equation}
 B^{1/2} \sim \bar \lambda_{f_R}  H \,,
 \end{equation}
 such that $B^{1/2}$ is essentially
 the Compton wavelength of $f_R$ at the background curvature
 in units of the horizon length.
 The Compton wavelength plays an important role in both cosmological
 and local tests of $f(R)$ models, as we shall see.

\begin{figure}[tb]\begin{centering}
\includegraphics[width=0.9\columnwidth]{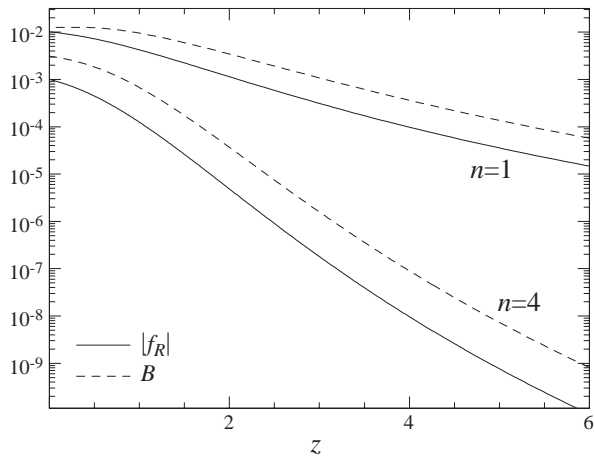}
\caption{Cosmological evolution of the scalar field $f_R$ and the Compton wavelength
parameter $B$ for models with $n=1,4$.   Both parameters control observable deviations
from general relativity and deviations decline rapidly with redshift as $n$ increases.}
\label{fig:fR}
\end{centering}\end{figure}

\subsection{Expansion History}
\label{sec:expansion}

We now evaluate the expansion histories for  the class of $f(R)$ models in Eqn.~(\ref{eqn:model}). First, we would like to narrow the parameter choices to yield expansion histories that are observationally viable, i.e.\ that deviate from $\Lambda$CDM in the effective equation of state (\ref{eqn:weff}) by no more than $|1+w_{\rm eff}| \la 0.2$ during the acceleration epoch. This equates to choosing a value for the field at the present epoch $f_{R0} \equiv f_R(\ln a=0) \ll 1$ or, equivalently, $R_0 \gg m^2$.  In this case, the approximation of Eqn.~(\ref{eqn:taylor}) applies  for the whole past expansion history and the field is always near the minimum of the effective potential
 \begin{equation}
 R = \kappa^2 \rho - 2f \approx \kappa^2 \rho + 2{c_1 \over c_2} m^2 \,,
 \end{equation}
 where the $2f$ term is nearly constant and mimics the energy density of  a cosmological constant.    Thus, to approximate
  the expansion history of $\Lambda$CDM with a cosmological constant
  $\tilde \Omega_\Lambda$ and matter density $\tilde \Omega_m$ with respect
  to a fiducial critical value, we set
 \begin{equation}
{c_{1} \over c_{2} }\approx 6{\tilde \Omega_{\Lambda}\over \tilde\Omega_{m}} \,,
\end{equation}
leaving two remaining parameters, $n$ and $c_1/c_2^2=6\tilde\Omega_\Lambda/c_2\tilde\Omega_m$ to control
how closely the model mimics $\Lambda$CDM.   Larger $n$ mimics $\Lambda$CDM until later in the expansion history; smaller $c_1/c_2^2$ mimics it more closely. Note that, since the critical
density and Hubble parameter depend on the $f_R$ modification,
 $\tilde \Omega_m$ is only the true value in the limit
\begin{equation}
\lim_{c_1 /c_2^2\rightarrow 0} \tilde\Omega_m = \Omega_m\,,
\end{equation}
whereas the matter density in physical units remains unchanged
$\tilde \Omega_m \tilde H_0^2 = \Omega_m H_0^2$.

For the flat $\Lambda$CDM expansion history
\begin{equation}
R \approx 3 m^2 \left( a^{-3} + 4{\tilde\Omega_\Lambda \over \tilde\Omega_m} \right) \,,
\end{equation}
and the field takes on a value of
 \begin{eqnarray}
f_R &=& -n{c_{1} \over c_{2}^{2}} \left( {m^{2}\over R} \right)^{n+1}  \,.
\label{eqn:minimum}
\end{eqnarray}
At the present epoch
\begin{eqnarray}
R_0 &\approx& m^2 \left( {12 \over\tilde \Omega_m} -9 \right)\,, \nonumber\\
f_{R0} &\approx &-n{c_{1} \over c_{2}^{2}} \left( {12 \over \tilde\Omega_m} -9 \right)^{-n-1} \,,\nonumber\\
B_0& \approx & {6n(n+1)\over {(1 + f_{R0})\tilde\Omega_m } }  {c_{1} \over c_{2}^{2}} \left( {12 \over \tilde \Omega_m} -9 \right)^{-n-2} \,.
\end{eqnarray}
In particular, for $\tilde \Omega_m = 0.24$ and $\tilde \Omega_\Lambda= 0.76$, $R_0 =
41 m^2$, $f_{R0} \approx -n c_1/c_2^2 /(41)^{n+1}$ and $B_0 \approx -0.61(n+1) f_{R0}$ for
$|f_{R0}| \ll 1$.  The consequences of cosmological and solar system-tests can be phrased in
a nearly model-independent way by quoting the field value $f_R$.
Consequently,  we will hereafter parameterize the amplitude $c_1/c_2^2$ through the cosmological field value today, $f_{R0}$.

In Fig.~\ref{fig:fR}, we show several examples of the background evolution of $f_R$.  For
a fixed present value $f_{R0}$, a larger $n$ produces a stronger suppression of the
field at high redshift and a larger value of $B$ relative to $f_{R}$.   The steepness of this
suppression will play an important role for galactic tests in \S \ref{sec:galaxy}.

\begin{figure}[tb]\begin{centering}
\includegraphics[width=0.9\columnwidth]{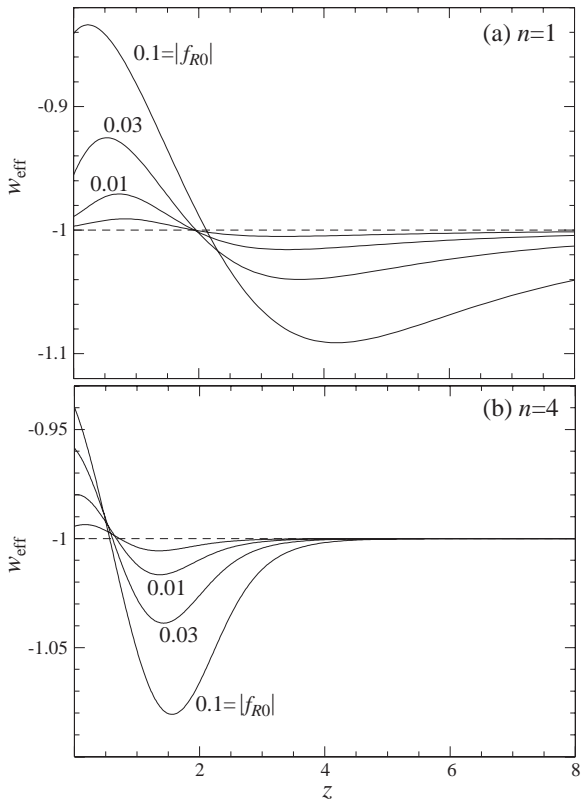}
\caption{Evolution of the effective equation of state for $n=1, 4$ for several values of the
cosmological field amplitude today, $f_{R0}$.  The effective equation of state crosses
the phantom divide $w_{\rm eff}=-1$ at a redshift that decreases with increasing $n$ leading
potentially to a relatively unique observational signature of these models.}
\label{fig:weff}
\end{centering}\end{figure}

The effective equations of state for these models are shown in Fig.~\ref{fig:weff}.  Deviations
from a cosmological constant, $w_{\rm eff}=-1$, are of the same order of magnitude as $f_{R0}$.
This class of models has a phantom effective equation of state, $w_{\rm eff}  < -1$, at high
redshift and crosses the phantom divide at a redshift that decreases with increasing $n$.
Note that an effective equation of state that evolves
across the phantom divide is a smoking gun for modified gravity acceleration
or dark energy with non-canonical degrees of freedom or couplings \cite{Vik04,Hu04c,Guoetal04,Amendola:2007nt}.

\subsection{Linear Perturbations}
\label{sec:linear}

Given an expansion history that defines $f_R(\ln a)$ and $B(\ln a)$,
the evolution of linear perturbations can be solved
using the techniques of \cite{SonHuSaw06}.  The principal feature of the linear evolution is
that once the wavelength of the perturbation becomes smaller than the Compton wavelength
in the background
\begin{equation}
{k \over a H} B^{1/2} > 1 \,,
\end{equation}
strong deviations from the GR growth rate appear.  In particular, the space-space $\Phi$ and time-time $\Psi$
pieces of the metric fluctuations in the Newtonian (longitudinal) gauge evolve to a ratio
\begin{equation}
-{\Phi \over \Psi} \equiv \gamma = {1 \over 2} \,,
\end{equation}
implying the presence of order-unity deviations from GR.

The consequence of this relative enhancement of
the gravitational potential $\Psi$ is an increase in the growth rate of linear density perturbations
on scales below the Compton wavelength.  If the Compton wavelength is longer than
the non-linear scale of a few Mpc, this transition leads to a strong and potentially observable
deviation in the matter power spectrum.    Even percent-level deviations in
the power spectrum are in principle detectable with future weak-lensing surveys.
If the Compton wavelength approaches the horizon, it can substantially alter the CMB
power spectrum as well \cite{SonHuSaw06}.

\begin{figure}[tb]\begin{centering}
\includegraphics[width=0.9\columnwidth]{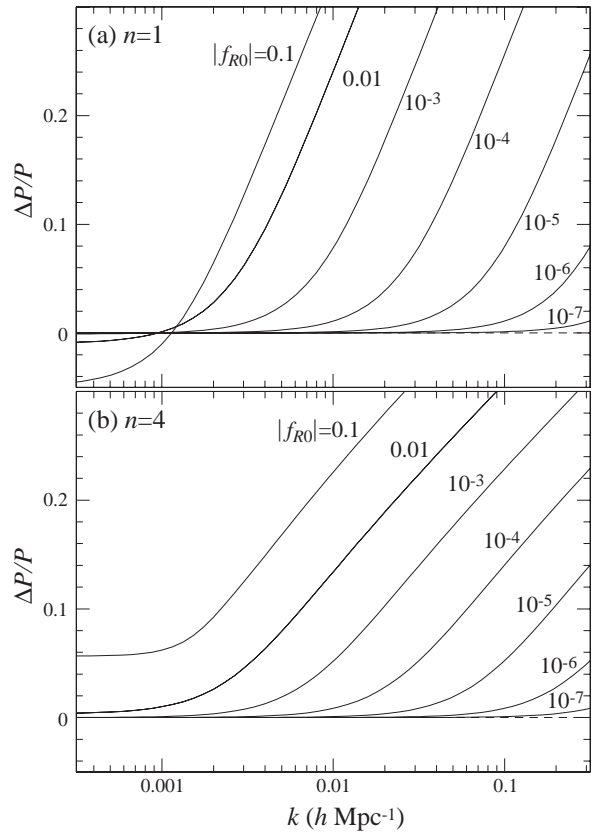}
\caption{Fractional change in the matter power spectrum $P(k)$ relative
to $\Lambda$CDM for a series of the
cosmological field amplitude today, $f_{R0}$, for $n=1,4$ models.  For scales that are below
the cosmological Compton wavelength during the acceleration epoch $k \ga (aH) B^{1/2}$
perturbation dynamics transition to the low-curvature regime where $\gamma=1/2$ and density
growth is enhanced.   This transition occurs in the linear regime out to field amplitudes of
$|f_{R0}| \sim 10^{-6}-10^{-7}$.}
\label{fig:Pk}
\end{centering}\end{figure}

In Fig.~\ref{fig:Pk}, we illustrate this effect for $n=1$ and $n=4$ models.   Deviations occur in the linear regime down to a field amplitude of $|f_R| \sim 10^{-7}$.  For these small field amplitudes, the expansion history and hence distance measures of the acceleration are indistinguishable from a cosmological constant with any conceivable
observational probe.  Nonetheless, linear structure can provide a precision test of gravity that, we shall see, rivals that of local tests in a substantially different curvature
regime.

\section{Local Tests}
\label{sec:local}

In this section, we consider local tests of $f(R)$ gravity.   We begin in \S \ref{sec:metric} with a general metric around spherically symmetric sources and its relationship to the $f_R$ field.
In \S \ref{sec:compton}, we discuss the qualitative behavior of the field
solutions and their relationship with the Compton
wavelength.   We evaluate solar-system constraints in \S \ref{sec:solar} and the requirements they place on the extent and evolution of the galactic halo in \S \ref{sec:galaxy}.

\subsection{Metric and Field Equations}
\label{sec:metric}

We take the general spherically symmetric isotropic form for the metric around a source centered at $r=0$
\begin{eqnarray}
    \ds s^2 &=& -[1-2\mA(r)+2\mB(r)] \ds  t^2  \nonumber\\\
    && + [1+2\mA(r)](\ds r^2 + r^2\ds\Omega) \,,
\end{eqnarray}
where we assume that $|\mA(r)| \ll 1$ and $|\mB(r)|\ll 1$ near the source, such that the metric is nearly Minkowski.  In the GR limit,  $\mB(r) \rightarrow 0$; limits on $\mB$ in the solar system
provide the strongest tests of modification to gravity of the type considered here.

By definition of the Ricci tensor, the sources of the metric potentials $\mA$ and $\mB$ are
given by
\begin{align}
    \nabla^2 (\mA+\mB) &= -\frac{1}{2}R \,, \\
    \nabla^2 \mB &= -\frac{1}{2}(R^{0}_{\hphantom{0}0}+R/2)
    \label{eqn:Bmetric} \,.
\end{align}
Note that a low-curvature $R \ll \kappa^2\rho$ solution may be accommodated by $\mB
\approx -\mA$.

To relate the time-time component of the Ricci tensor to $f(R)$, let us take the time-time component of the field equation (\ref{eqn:metricvar})
\begin{equation}
(1+f_R) R^{0}_{\hphantom{0}0} - {1\over 2} (R+f) + \Box f_R + \partial_t^2 f_R = -\kappa^2 \rho \,.
\end{equation}
Under the assumption of a static solution, we can combine this equation with the trace equation
(\ref{eqn:trace}) to obtain
\begin{eqnarray}
R^{0}_{\hphantom{0}0} = {-2 \kappa^2 \rho + {1\over 2} R - {1\over 2} f + f_R R \over 3(1+f_R)}\,.
\end{eqnarray}
Even in a low-curvature solution where
$R \ll \kappa^2 \rho$, $R^{0}_{\hphantom{0}0} = {\cal O}(\kappa^2\rho)$.

Eqn.~(\ref{eqn:Bmetric}) then becomes
\begin{eqnarray}
\nabla^2 \mB  &=& -{1 \over 4} \left( { -4 \kappa^2 \rho + 4 R + 5 f_R R - f \over 3(1+f_R)} \right)\,,
\label{eqn:gammasource}
\end{eqnarray}
where $f(R)$ is given by the solution to the trace equation (\ref{eqn:trace}) in the static limit
 \begin{eqnarray}
 3\nabla^2 f_R - R + f_R R - 2 f &=& -\kappa^2 \rho  \,.
 \label{eqn:statictrace}
\end{eqnarray}

As an aside, choosing models for which $B>0$ in \S\ref{sec:model}, not only results in the existence in the expansion history of a stable matter-dominated era, but ensures that the models do not exhibit the related instability for stellar-type objects \cite{DolKaw03,Sei07}.
Small, time-dependent perturbations to a high-curvature solution of Eqn.~\eqref{eqn:statictrace} have positive mass squared and do not grow in this class of $f(R)$.

In the limit that $|f_R | \ll 1$ and $|f/R| \ll 1$, valid for all sources that we
shall consider,
\begin{eqnarray}
\nabla^2 \mA & \approx & -{1\over 2}\kappa^2\rho + {1 \over 6} \left( \kappa^2 \rho - R \right)\,,
\label{eqn:Apoisson}\\
\nabla^2 \mB  & \approx & {1\over 3}\left( \kappa^2 \rho - R \right)
\label{eqn:Bpoisson} \,.
\end{eqnarray}
As expected, the source of $\mB$ is the deviation of the curvature $R$ from the GR value of $\kappa^2\rho$.
Moreover, in the same limit, equation~(\ref{eqn:statictrace}) for $f_R$ becomes
\begin{equation}
\nabla^2 f_R \approx {1\over 3} \left( R - \kappa^2 \rho \right) \,.
\label{eqn:statictracelim}
\end{equation}
A solution for $f_R$ therefore gives $\mB$ up to constants of integration
\begin{equation}
\mB(r) = -f_R(r) + a_1  +  {a_2 \over r} \,.
\end{equation}
Since $\mB$ must remain finite at $r=0$, $a_2=0$.  Let us assume that
at sufficiently large radii $f_R(r) \rightarrow f_{R\infty}$ and
$\mB \rightarrow 0$ then
\begin{equation}
\mB(r) = - [f_R(r) - f_{R\infty}] \equiv -\Delta f_R(r) \,.
\end{equation}
For radii beyond which the source
$\kappa^2\rho-R$ becomes negligible $\mB(r) \propto 1/r$.   It is convenient to then define an effective enclosed mass
\begin{equation}
M_{\rm eff}  = 4\pi \int (\rho-R/\kappa^2) r^2 \ds r \,,
\end{equation}
such that
\begin{equation}
\mB(r) = -\Delta f_R(r) \rightarrow { 2 G M_{\rm eff} \over 3 r} \,.
\label{eqn:deltaf}
\end{equation}
Note that this is an implicit solution, since $R(f_R)$.  Nevertheless, we shall see in the next section that it sheds light on the behavior of explicit solutions.

Finally, the deviation from the general-relativistic metric is given by
\begin{equation}
\gamma- 1\equiv {\mB \over \mA-\mB}  \rightarrow -{2M_{\rm eff}\over 3 M_{\rm tot}+ M_{\rm eff}}\,,
\label{eqn:gammaexterior}
\end{equation}
where $M_{\rm tot}$ is the total mass of the system.    The two limiting cases are
$M_{\rm eff} \ll M_{\rm tot}$ for which $\gamma - 1 = -2M_{\rm eff}/3M_{\rm tot}$ and
$M_{\rm eff} = M_{\rm tot}$ for which $\gamma-1 = -1/2$.  These solutions correspond to
high curvature, $R \approx \kappa^2\rho$, and low curvature, $R \ll \kappa^2 \rho$
\cite{EriSmiKam06,Chiba:2006jp} respectively.

\subsection{Compton and Thin-Shell Conditions}
\label{sec:compton}

Before examining explicit solutions for $f_{R}(r)$ given $\rho(r)$, we show how the nature of the solutions is tied to the Compton wavelength of the field and exhibits the so-called chameleon mechanism \cite{KhoWel04} for hiding scalar degrees
of freedom in high-density regions \cite{NavAco06,FauTegBunMao06}.

An examination of Eqn.~(\ref{eqn:statictrace}) shows that there are two types of {\it local} solutions to the field equations distinguished by a comparison of the Compton wavelength of the field $f_R$
\begin{equation}
\lambda_{f_R} \equiv m_{f_R}^{-1} \approx \sqrt{3 f_{RR}} \,,
\end{equation}
to the density structure of the source.

Let us again assume that $|f_R|\ll1$ and $|f/R|\ll 1$ so that the field equation is well approximated by Eqn.~(\ref{eqn:statictracelim}).
The first class of solutions to this equation has high curvature $R \approx \kappa^2\rho$ and
small field gradients
$\nabla^2 f_R \ll \kappa^2 \rho$.  The
second class of solutions has low curvature
 $R \ll \kappa^2\rho$ and large field gradients $\nabla^2 f_R \approx - \kappa^2 \rho/3$.

 A {\it sufficient} condition for the high-curvature solution is that field gradients
 can be ignored at all radii when compared with the density source.
A {\it necessary} or consistency condition is that
field gradients implied by the high-curvature solution $f(R=\kappa^2 \rho)$
can be ignored compared with local density gradients.   More specifically
\begin{equation}
f_{RR}\Big|_{R=\kappa^2\rho}
 \partial_i^2 \rho \ll \rho\,, \qquad f_{RR}^{1/2} \Big|_{R=\kappa^2\rho} \partial_i \rho \ll \rho \,,
\label{eqn:comptoncondition}
\end{equation}
i.e.\ that the density changes on scales that are much longer than the Compton wavelength.
Equivalently, a mass source induces changes in the field with a Yukawa profile of
$e^{-m_{f_R} r}/r$ which are highly suppressed on scales larger than the Compton wavelength.
We call this condition the
Compton condition.

If the Compton condition is satisfied at all radii, then the high-curvature solution is also valid
at all radii and deviations from GR will be highly suppressed.   If this condition is violated
beyond some outer radius, then a portion of the
exterior must be at low curvature  $R \ll \kappa^2 \rho$.
Moreover, since Birkhoff's theorem does not apply,
the exterior low-curvature solution can penetrate into the region where
the Compton condition~(\ref{eqn:comptoncondition}) is locally satisfied.
The interior solution then depends
on the exterior conditions.  We have seen in \S \ref{sec:linear} that linear cosmological
perturbations are in the low-curvature regime on scales smaller than the cosmological
Compton wavelength and so the Compton criteria must be violated far in the exterior
if $|f_{R0}| \ga 10^{-7}$.  We will return to this point in \S \ref{sec:galaxy}.

To quantify these considerations, note that the maximal change in $f_R$ from the
interior to the exterior is imposed by the low-curvature assumption $R \ll \kappa^2\rho$ or
$M_{\rm eff} = M_{\rm tot}$
\begin{equation}
\Delta f_R(r) \le {2 \over 3} \Phi_M(r) \,,
\label{eqn:globalthinshell}
\end{equation}
where $\Phi_M(r)$ is the Newtonian potential profile of the source, i.e. $\Phi_M \approx
 G M_{\rm tot}/r$
exterior to the dominant mass.
This condition sets an upper limit
on the difference between the interior and exterior field values for a static solution.

If the thin-shell condition is satisfied and $|\Delta f_R(r)| \ll \Phi_M(r)$, then $M \gg M_{\rm eff}$ and somewhere in the interior there must exist a high-curvature region where $R \rightarrow \kappa^2\rho$.  To estimate where this occurs consider a local version of  Eqn.~(\ref{eqn:globalthinshell})
\begin{equation}
\Delta f_R \la  \kappa^{2} (\rho-\rho_{\infty}) r^{2} \,.
\label{eqn:localthinshell}
\end{equation}
From the outside in, when this condition is first satisfied, there is enough source to make the transition between the interior and exterior field values. Once this is satisfied, it remains satisfied in the interior as long as further changes in $f_R$
are much smaller than the initial jump. In other words, the exterior field is only generated by the ``thin shell" of mass $M_{\rm eff}$ that lies outside of this transition.   We will call this the thin-shell criterion and such a solution is known in the literature as a chameleon solution.

The thin-shell criterion is related to the low-curvature linearization condition of
 \cite{Chiba:2006jp} Eqn.\ (13).   There, a solution for the curvature is found by linearizing Eqn.~\eqref{eqn:statictrace} around its background value. Requiring that the value of the perturbation be smaller than that of the background curvature, results in exactly the opposite of Eqn.~\eqref{eqn:globalthinshell}
 \begin{equation}
 |f_{R\infty}| \ga  {2 \over 3} \Phi_M(r) \,.
 \end{equation}
 Therefore, the linearization procedure is not valid for exactly those sets of parameters for which the thin-shell condition is satisfied: high-curvature solutions are necessarily non-linear. When the linearization is valid, the solution is low-curvature everywhere leading to large deviations from
 GR in the interior.

The thin-shell criterion~(\ref{eqn:localthinshell}) is also related to but stronger than
the local Compton condition
\begin{equation}
f_{RR} < r_\rho^2 \,,
\end{equation}
where $r_\rho$ is the distance over which the density field changes.   Converting
derivatives to finite differences $f_{RR} \approx \delta f_R/\delta R$
\begin{equation}
\delta f_R \la \delta R r_\rho^2 \approx \kappa^2 \delta \rho r_\rho^2 \,.
\label{eqn:comptoncondition2}
\end{equation}
The difference between the two conditions is that the Compton condition involves
the small change in the field $\delta f_R$ at high curvature, whereas the thin shell criterion
involves the potentially larger change in the field $\Delta f_R$ from the high- to low-curvature regimes.
If there is no transition to low curvature in the exterior then these conditions are the same.

The thin-shell condition implies that the field does not always sit at the local potential minimum $R \approx \kappa^2 \rho$.   Nonetheless, the field does choose the  energetically favorable configuration.  The field $f_{R}$ will not lie at the potential minimum in the interior region if the energy cost for introducing a field gradient between the
interior and exterior is too high. The potential energy density
cost for not lying at the potential minimum is
\begin{equation}
\Delta V \sim {\kappa^{2} \over 3} \rho |\Delta f_{R} | \,,
\end{equation}
where $\Delta f_{R}$ is the difference between potential minimum and the exterior solution. Compare this with the gradient energy density gain from not introducing a profile in the field
\begin{equation}
\Delta E \sim {1\over 2} {| \Delta f_{R}|^{2} \over r^{2}} \,.
\end{equation}
The potential energy cost outweighs the gain if
\begin{equation}
|\Delta f_{R}| \la  \kappa^{2}\rho r^{2} \,,
\end{equation}
which is the thin-shell condition.

\begin{figure}[tb]\begin{centering}
\includegraphics[width=0.9\columnwidth]{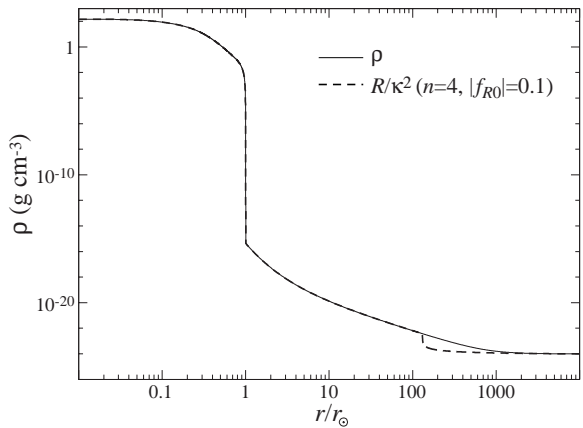}
\caption{Density profile in the solar interior and vicinity (solid curve).  Under general
relativity (GR), the curvature $R$ would track the density profile.  For the $f(R)$ model
with $n=4$, cosmological field amplitude $|f_{R0}|=0.1$ and a galactic field amplitude
that minimizes the scalar potential, the curvature tracks the GR or high-curvature limit
out to the edge of the solar corona at about 1AU (dashed line). }
\label{fig:rho}
\end{centering}\end{figure}

\subsection{Solar System to Galaxy}
\label{sec:solar}

We now consider explicit solutions of the field equation~(\ref{eqn:statictrace}),
given the density profile $\rho(r)$, in the solar vicinity.
In this section, we will assume that the galaxy has sufficient mass and extent
to bring the field to the potential minimum of  Eqn.~(\ref{eqn:minimum}) in
the outskirts of the solar system.   In the
next section, we will discuss the requirement this assumption places on the
 structure and
evolution of the galactic halo.

Specifically, we set the boundary condition
$f_{R\infty}= f_{Rg}$ where
\begin{equation}
f_{Rg}=f_R(R=\kappa^2\rho_g) \,.
\label{eqn:fRg}
\end{equation}
Here, $\rho_g$ is the average galactic density in the solar vicinity.
Under this assumption, the galactic field value is related to the cosmological
one as
\begin{equation}
\left( f_{Rg} \over f_{R0} \right)^{1 \over n+1} = 8.14\times 10^{-7}
{R_0 \over m^2} {\Omega_m h^2\over 0.13}
\left( {\rho_g \over 10^{-24} {\rm g\ cm}^{-3}} \right)^{-1} \,.
\label{eqn:fRgfR0}
\end{equation}

For the density profile in the solar vicinity we take the helioseismological
model of \cite{BahSerBas05} for the solar interior, the chromosphere model of
\cite{VerAvrLoe81} and the corona model of \cite{SitGuh99} for the surrounding
 region.   We add to these
 density profiles a constant galactic density of $\rho_g =10^{-24}$ g cm$^{-3}$.
 This profile is shown in
 Fig.~\ref{fig:rho} (solid curve).

We solve the field equation \eqref{eqn:statictrace} as a boundary-value problem using a
relaxation algorithm.  The source-free $(\rho=0)$ field equation has exponentially growing
and decaying Yukawa homogeneous solutions $\exp(\pm m_{f_R}r)/r$.   Initial-value integrators
have numerical errors that would stimulate the positive exponential, whereas relaxation
methods avoid this problem by enforcing the outer boundary at every step.
We show an example solution in Fig.~\ref{fig:rho} (dashed curve) for $n=4$ and $|f_{R0}|=0.1$.

\begin{figure}[tb]\begin{centering}
\includegraphics[width=0.9\columnwidth]{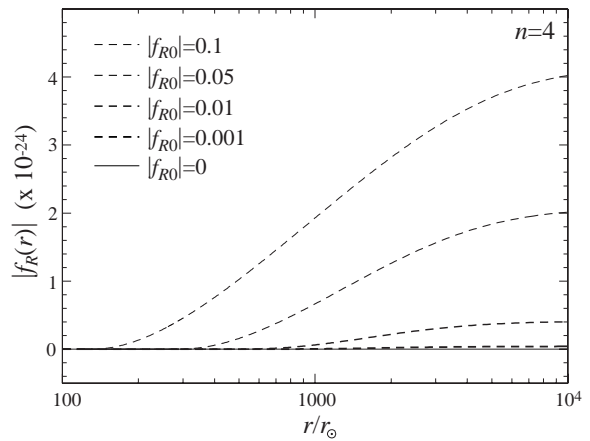}
\caption{Solar solution for $f_R(r)$ with $n=4$ for a series of cosmological field amplitudes
$f_{R0}$ with a galactic field that minimizes the potential.
The field smoothly transitions from the $f_R \sim 0$ interior value to the galactic values
once the thin-shell criterion is violated.}
\label{fig:fRsolar}
\end{centering}\end{figure}

A solution is found on an interval, by requiring that the field $f_R$ minimize the potential
\eqref{eqn:potn} at two chosen radii: one far away from the Sun and its
corona, the other---inside the solar density distribution. We place
the outer boundary at
$r = 10^6 r_\odot$, where the density distribution is entirely dominated by the constant
galactic-density component. The solutions are robust to increasing the radius
of this boundary.

For the inner boundary, we take the starting point as approximately 1000 Compton wavelengths
from the transition to low curvature; the Compton condition is well satisfied there.
Interior to this point, the solution is more efficiently obtained
by a perturbative solution around the high-curvature solution of $R^{(0)}=\kappa^{2}\rho$.
The first-order correction to this solution is $R= R^{(0)} + R^{(1)}$ with
 \begin{equation}
 R^{(1)} = \left( 3 \nabla^2 f_R + f_R R - 2f \right)\Big|_{R=R^{(0)}} \,.
\end{equation}
With this correction to $R$, the first-order correction to $f_{R}$ can be obtained
using  initial-value-problem methods.   This series can be iterated to arbitrary
order and is strongly convergent when the Compton condition is satisfied.
In fact, at our chosen starting point, the zeroth order solution typically suffices.
We have also checked that for cases where
the Compton condition is everywhere satisfied, the numerical solution matches
the perturbative solution.
Finally, we stop the relaxation code when the solution satisfies the
  trace equation \eqref{eqn:statictrace} to at
least $10^{-6}$ accuracy.

Let us now relate the numerical solutions shown in Fig.~\ref{fig:fRsolar}
 to the qualitative analysis of the
previous section.
First,  consider the Compton condition.
The Compton wavelength at $R= \kappa^2\rho$ is
\begin{eqnarray}
\lambda_{f_R} &=& (10.6 {\rm pc}) \left( 8.14\times 10^{-7} \right)^{(n-1)/2}
 \nonumber\\
 && \times [(n+1)|f_{R0}|]^{1/2}  \left({ R_0 \over m^2}{\Omega_m h^2 \over 0.13}  \right)^{(n+1)/2} \nonumber\\
 && \times \left( {\rho \over 10^{-24} {\rm g\ cm}^{-3}} \right)^{-(n+2)/2}\,.
 \label{eqn:comptongalactic}
\end{eqnarray}
For example, for $n=4$ and the fiducial cosmology
\begin{equation}
\lambda_{f_R} \approx  (8300r_\odot) |f_{R0}|^{1/2}
\left( {\rho \over 10^{-24} {\rm g\ cm}^{-3}}  \right)^{-3} \,,
\end{equation}
so that the Compton condition is satisfied for the whole solar profile for $|f_{R0}| \la 10^{-2}$.
Note that, even at the base of the corona where the density is $\rho \approx 10^{-15}$ g cm$^{-3}$,
the Compton wavelength is $\sim 10^{-23}|f_{R0}|^{1/2} r_\odot$.   Thus,
in spite of the steep density gradient at the edge of the Sun, the Compton condition is
well satisfied, allowing the solution in Fig.~\ref{fig:rho} to follow the high-curvature
solution $R=\kappa^{2}\rho$ closely through the transition.

\begin{figure}[tb]\begin{centering}
\includegraphics[width=0.9\columnwidth]{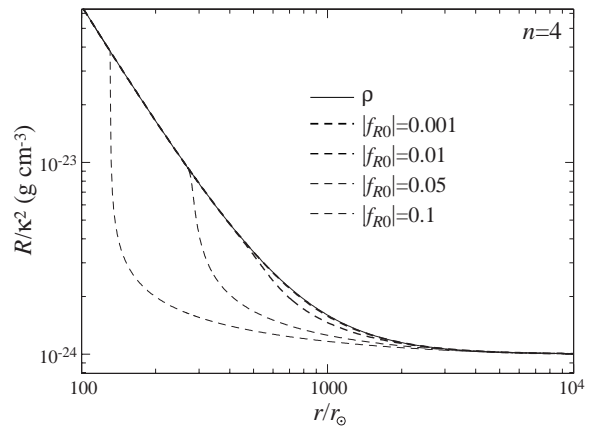}
\caption{Solar solution for the curvature $R$ for the $n=4$ model and  a series of cosmological field amplitudes
$f_{R0}$ with a galactic field that minimizes the potential.  The  solution abruptly
transitions from the high-curvature
$R \approx \kappa^2 \rho$ to the low-curvature regime once the thin-shell criterion is
violated.}
\label{fig:Rsolar}
\end{centering}\end{figure}

On the other hand, the thin-shell criterion is satisfied in the
solar corona up to $|f_{R0}| \la 10^{-1}$.  Thus for $n=4$,
we expect order-unity cosmological fields to be achievable with the entire solar interior
including the edge and chromosphere in the high-curvature regime.

The numerical solutions shown in Fig.~\ref{fig:fRsolar} verify the qualitative behavior
described in the previous section.   For $|f_{R0}| \la 10^{-2}$, the deviations from
the high-curvature $R=\kappa^2\rho$ limit are fractionally small since the Compton
condition is everywhere satisfied.  For $|f_{R0}| \la 10^{-1}$, the break to low curvature
occurs in the corona.   This break occurs gradually in the field profile $f_R(r)$ but
rapidly in the curvature (see Fig.~\ref{fig:Rsolar}).  At small field values, a small
change in $f_R$ represents a large change in the curvature.   $M_{\rm eff}$ is approximately
just the mass between this transition and the point at which the galactic density exceeds
the corona.  Outside of this transition, the exterior field
relaxes to the galactic value as $|\Delta f_R| \propto e^{-m_{f_R}r}/r$.
In these examples, the Compton wavelength in the galaxy is of order $10^3-10^4$ $r_\odot$ and
the mass term further suppresses the deviations from GR.

\begin{figure}[tb]\begin{centering}
\includegraphics[width=0.9\columnwidth]{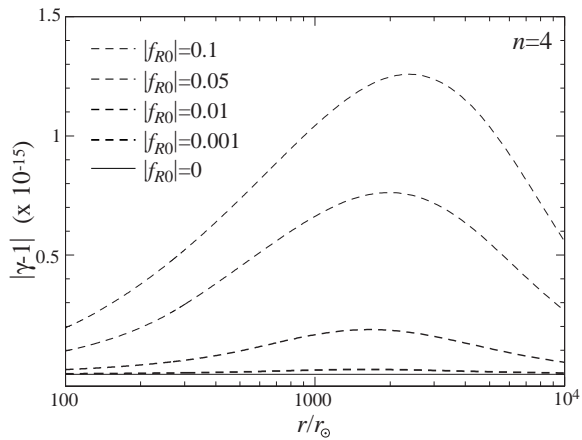}
\caption{Metric deviation parameter $|\gamma-1|$ for
$n=4$ models and  a series of cosmological field amplitudes
$f_{R0}$ with a galactic field that minimizes the potential.  These deviations
are unobservably small for the whole range of amplitudes.}
\label{fig:gamma}
\end{centering}\end{figure}

In Fig.~\ref{fig:gamma} we show $|\gamma-1|$ for the same $n=4$ models.   The deviations
peak at $\sim 10^{-15}$.
Such deviations easily pass the stringent solar system tests of gravity from the
Cassini mission \cite{Will:2005va}
\begin{equation}
|\gamma -1| < 2.3 \times 10^{-5}
\end{equation}
under the assumption that
the galactic field $f_{Rg}$ is given by the potential
minimum.

Models that saturate this observational bound
 have a sufficiently large $|f_{Rg}|$ that the thin-shell
criterion is first satisfied at the edge of the Sun, where the fractional
enclosed mass becomes $2M_{\rm eff}/3 M_{\rm tot} \approx 10^{-5}$
(see Eqn.~\ref{eqn:gammaexterior}). For a given $f_{R0}$, this can be achieved
by lowering $n$.   In such models, the additional contribution to the
effective mass outside of the photosphere
is negligible and the field solution obeys Eqn.~(\ref{eqn:deltaf}). The
exterior solution therefore
becomes
\begin{equation}
\Delta f_R(r)  \approx (\gamma-1)\Phi_M(r) \approx (\gamma-1) {G M_\odot \over r} \,,
\end{equation}
since the Compton wavelength in the exterior implied by
Eqn.~(\ref{eqn:comptongalactic})  is much larger than the solar system.

\begin{figure}[tb]\begin{centering}
\includegraphics[width=0.9\columnwidth]{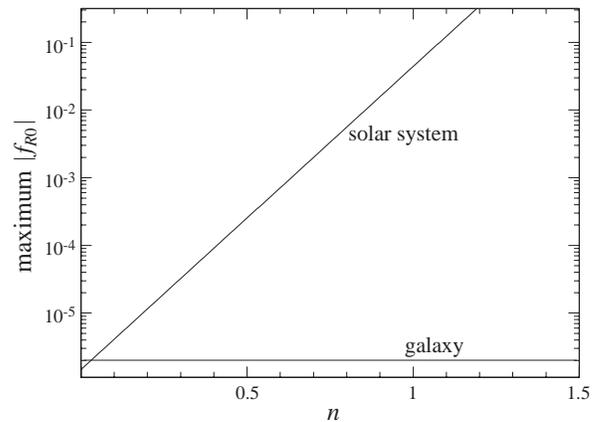}
\caption{Maximum cosmological field $|f_{R0}|$ allowed by the solar system constraint
alone (top curve)
under the  assumption that the galactic field remains at the potential minimum
at the present epoch.  Models with $|f_{R0}| \ga 10^{-6}$  (bottom curve) have
galactic fields $|f_{Rg}|$ that evolve to higher values during the acceleration epoch, potentially
enabling substantially stronger constraints with structure formation simulations.}
\label{fig:gammalim}
\end{centering}\end{figure}

Given that
\begin{equation}
{G M_\odot \over r_\odot} = 2.12 \times 10^{-6} \,,
\end{equation}
the solar-system constraints can be simply stated as
\begin{equation}
|\Delta f_R(r_\odot)| \approx | f_{Rg} |  <  4.9 \times 10^{-11} \,.
\label{eqn:fRgconstraint}
\end{equation}
Note that this bound is independent of the form of $f(R)$ and the assumption that
$f_{Rg}$ is given by the minimum of the effective potential.
The model dependence comes from the implications for the cosmological field value
$f_{R0}$. Using Eqn.~(\ref{eqn:fRgfR0}) at a galactic density of $\rho_g = 10^{-24}$~g~cm$^{-3}$,
 we can translate this into a bound on the amplitude of the
cosmological field
\begin{equation}
|f_{R0}| < 74  \left(  1.23 \times 10^{6} \right)^{n-1}
\left[ {R_0 \over m^2}
{\Omega_m h^2 \over 0.13}\right]^{-(n+1)} \,.
\label{eqn:fR0constraint}
\end{equation}
As shown in Fig.~\ref{fig:gammalim} this is a fairly weak constraint.  For $n>1$ it allows order
unity cosmological deviations from GR.  Note that the models of \cite{FauTegBunMao06}
are equivalent to a continuation of the approximate form of our model
in Eqn.~(\ref{eqn:taylor}) but with
$n < 0$.  Solar system constraints on cosmological amplitudes are significantly stronger
in that class.  Finally, although a detailed calculation is beyond the scope of this work,
laboratory constraints on fifth forces are also weak under the same assumptions
given the much larger effective densities involved \cite{KhoWel04}.

\subsection{Galaxy to Cosmology}
\label{sec:galaxy}

The cosmological bound Eqn.~(\ref{eqn:fR0constraint})  from the
solar-system constraint Eqn.~(\ref{eqn:fRgconstraint})
 is robust but weak.   A related but potentially more powerful constraint comes from
 the transition between the high curvature of the galaxy and the desired
 cosmological curvature.
  Indeed, the implicit assumption in applying
 the solar-system constraint to the cosmology is that
 the galaxy itself is in the high-curvature regime with respect to its own density profile.

The validity of this assumption
depends on both the structure of the galactic halo and its evolution during the acceleration epoch.
 Large $|f_{R0}|$ requires that either the galactic gravitational potential is substantially deeper than
in $\Lambda$CDM or that it has not yet reached its equilibrium value.

\begin{figure}[tb]\begin{centering}
\includegraphics[width=0.9\columnwidth]{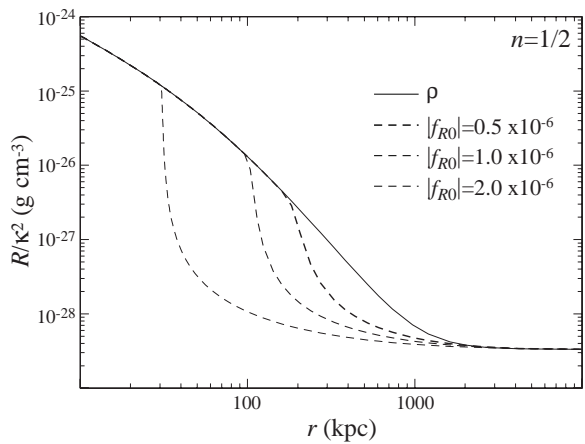}
\caption{A galactic solution for the scalar curvature $R$ for $n=1/2$  with an
NFW density profile (\ref{eqn:nfw}) and conservative parameters
$r_s = 75$~kpc and $v_\mathrm{max} = 300$~km s$^{-1}$. For $|f_{R0}| \ga 2\times 10^{-6}$, no \emph{static} solution can be found which has high curvature, $R \approx \kappa^2\rho$, inside the halo.  As in the solar case,
once the thin-shell criterion is violated the solution abruptly transitions
to low curvature and matches onto the cosmological value of $f_{R0}$ outside the halo.}
\label{fig:haloR}
\end{centering}\end{figure}

  From the linear theory analysis in \S \ref{sec:linear}, we know that the Compton condition is
  violated for linear perturbations if $|f_{R0}| \ga 10^{-7}$.   Hence, at the outskirts of the galactic
  halo where the density profile joins onto the large-scale structure of the universe, there
  must be a transition from high to low curvature.
As we have seen in \S \ref{sec:compton}, the low-curvature cosmological field will eventually penetrate into
the galaxy unless
\begin{equation}
|\Delta f_R(r)|  \equiv |f_{Rg}-f_{R0}| \approx |f_{R0}| \la {2 \over 3}\Phi_g \,,
\end{equation}
where $\Phi_g$ is the Newtonian potential of the galaxy.    For definiteness, let us consider
the NFW density profile
\begin{equation}
\rho_g(r) = {M_g \over 4\pi}{1 \over r(r+r_s)^2} \,, \label{eqn:nfw}
\end{equation}
where $M_g$ is the galactic mass contained within $5.3 r_s$ and $r_s$ is the scale
radius of the dark-matter halo.
This density profile has a Newtonian potential of
\begin{equation}
\Phi_g = { G M_g \over r}\ln (1+r/r_s)  \,,
\end{equation}
and a maximum rotation-curve velocity of
\begin{equation}
v_{\rm max} = 0.46 \left( { G M_g \over r_s} \right)^{1/2}
\end{equation}
at $2.16 r_s$.
Taking the thin-shell criterion to be satisfied at $r \approx r_s$, such that the interior
is in the high-curvature regime, leads to an upper limit on $|f_{R0}|$ for
a static solution of
\begin{equation}
 |f_{R0}|  \la 2 \times 10^{-6} \left( { v_{\rm max} \over 300{\rm km/s}} \right)^2 \,.
 \label{eqn:fR0bound}
\end{equation}
For higher cosmological values of the field $|f_{R0}|$, the galaxy will relax over time to the low-curvature solution to minimize the cost of field gradients.     Note that even a value of
$|f_{R0}| = 10^{-6}$ provides potentially observable modifications in the linear
regime since gravity at the juncture is still modified by order unity
(see Fig.~\ref{fig:Pk}).   We have verified through the numerical techniques of
the previous section that the constraint in Eqn.~(\ref{eqn:fR0bound}) is nearly independent
of the functional form of $f(R)$. In Fig.~\ref{fig:haloR}, we show an example
solution.  Here we err on the
conservative side by taking $v_{\rm max} = 300$~km s$^{-1}$ and $r_s=75$kpc to reflect a somewhat
more massive and extended halo by a factor of a few
 than expected in $\Lambda$CDM for our galaxy
 (c.f. \cite{KlyZhaSom02}, Fig. 7).  Empirical data exists only out to $\sim 20$kpc where the rotational velocity reaches $v \approx
230$~km s$^{-1}$.

Unfortunately, the bound in Eqn.~(\ref{eqn:fR0bound}) is suggestive but not definitive.
A cosmological simulation will be required to determine how
haloes of galactic size, which are embedded in a group-sized dark-matter halo, which itself
is part of the quasi-linear
large-scale structure of the universe,    evolve during the acceleration epoch.
The density profiles of the structures in which the galaxy is embedded can further shield
the galactic interior from the low-curvature solution.  Furthermore, the
cosmological background itself was at high curvature at $z\gg 1$.
For $n \gg 1$, the
condition that the background $|f_{R}| \la 10^{-6}$ is well satisfied for $z\gg 1$ even for $f_{R0}$
approaching unity.  For example if $n=4$, the cosmological field drops by $10^{3}$ by $z=2$
(see Fig.~\ref{fig:fR}).  Therefore when the galactic halo formed both its interior and exterior
were at high curvature.
The local curvature would then follow the local density $R \approx \kappa^2\rho$
closely everywhere.

Only during the recent acceleration epoch is the thin-shell criterion for the galaxy violated
at $|f_{R0}| \ga 10^{-6}$.  The low curvature, high field values of the background will then begin to propagate into the interior
of the galactic halo in a manner that requires a simulation of the process.
The static, thin-shell bound for the galaxy
given by Eqn.~(\ref{eqn:fR0bound}) is therefore
overly restrictive but suggests that cosmological simulations  should
enable much more stringent bounds on $f(R)$ models than solar-system test alone.

\section{Discussion}
\label{sec:discussion}

We have introduced a class of $f(R)$ models that accelerate the expansion without
a cosmological constant.  Its parameters allow the gravitational phenomenology
exhibited in cosmological, galactic and solar-system tests to span the range between infinitesimal
and order-unity deviations from general relativity with a cosmological constant.

In these models, unlike the original model of \cite{Caretal03} and related generalizations,
the general-relativistic or high-curvature value for  the Ricci scalar $R \approx \kappa^2\rho$
is the solution that minimizes the potential for the scalar degree of freedom
$f_R = \ds f(R)/\ds R$.
This feature is critical for both cosmological and solar-system tests.

Solar-system tests of $f(R)$ gravity {\it alone} place only weak bounds on these models, despite a strong and nearly model-independent limit on the $f_R$ field amplitude in the galaxy of $|f_{Rg}| < 5 \times 10^{-11}$.  Indeed, we show that for
a range of models that include cosmological fields of order
unity $|f_{R0}| \la 1$,  the field deviates from the high-curvature regime only
for a brief interval in the solar corona.    Likewise, deviations from general relativity in
the metric are generated by the mass in this interval and so are unobservably small.
This is strikingly different from constraints that would arise if one embeds the Sun directly
into a medium at the cosmological density $|f_{R0}| < 5 \times 10^{-11}$.

Without further constraints on the size and evolution of the galactic halo, solar-system tests
may be evaded relatively easily.
This is because of the assumed strong density scaling of the
field amplitude at the potential minimum.   However, though the galactic field begins at high
redshift at its potential minimum, it will only remain there if the galaxy is sufficiently massive
to protect it against the cosmological exterior which is evolving to low curvature at $z < 1$.

An order of magnitude estimate, based on the extrapolation of rotation-curve measurements
and the assumption that the galactic halo does not differ substantially from $\Lambda$CDM
expectations in the outskirts, suggests that an isolated galaxy that is otherwise like our own
will only remain stably at high curvature
if the cosmological field is $|f_{R0}| \la 10^{-6}$.   Despite the fact that the high-curvature
solution still minimizes the potential energy of the field, the gradient energy implied by
the field profile from the galactic interior to exterior is too high.
This estimate is also nearly independent
of the functional form of $f(R)$ and improvements in the solar-system constraint.
Turning this estimate into a firm constraint on models
will require cosmological simulations of $f(R)$ acceleration to examine how far into the solar system the exterior low-curvature field values can penetrate by the present epoch in the local environment of our galactic halo.

Distance-based measurements of the expansion history
will be limited to testing $|f_{R0}|\ga 10^{-3}-10^{-2}$ for the foreseeable future, since
the field amplitude determines the deviations in the effective equation of state to be of comparable size.
 Nonetheless, future, percent-level constraints on the
matter power spectrum in the linear regime offer potentially even stronger tests of $f(R)$ models than
solar or galactic constraints,
in principle down to amplitudes of $|f_{R0}| \sim 10^{-7}$.   This sensitivity is due to the large
Compton scale in the background across which perturbations make the transition from low
to high curvature and exhibit order-unity deviations from general relativity.  Cosmological simulations are also
required to determine how these signatures can be disentangled from the non-linear evolution of
structure.

\vspace{1.cm}

\noindent {\it Acknowledgments}: We thank S. Basu, S. Carroll,  S. DeDeo, A. Erickcek,
M. Kamionkowski,
D. Kapner, D. Psaltis, B. Robertson,
M. Seifert, T. Smith, A. Upadhye, B. Wald and A. Weltman for useful conversations.
This work was supported by the
U.S.~Dept. of Energy contract DE-FG02-90ER-40560. IS and WH are
additionally supported by the David and Lucile Packard Foundation
and by the KICP under NSF PHY-0114422. IS thanks for the hospitality of the California Institute of Technology where a part of this work was carried out.
\hfill\vfill
\bibliography{fr_solar}

\begin{thebibliography}{96}
\expandafter\ifx\csname natexlab\endcsname\relax\def\natexlab#1{#1}\fi
\expandafter\ifx\csname bibnamefont\endcsname\relax
  \def\bibnamefont#1{#1}\fi
\expandafter\ifx\csname bibfnamefont\endcsname\relax
  \def\bibfnamefont#1{#1}\fi
\expandafter\ifx\csname citenamefont\endcsname\relax
  \def\citenamefont#1{#1}\fi
\expandafter\ifx\csname url\endcsname\relax
  \def\url#1{\texttt{#1}}\fi
\expandafter\ifx\csname urlprefix\endcsname\relax\def\urlprefix{URL }\fi
\providecommand{\bibinfo}[2]{#2}
\providecommand{\eprint}[2][]{\url{#2}}

\bibitem[{\citenamefont{Carroll et~al.}(2004)\citenamefont{Carroll, Duvvuri,
  Trodden, and Turner}}]{Caretal03}
\bibinfo{author}{\bibfnamefont{S.~M.} \bibnamefont{Carroll}},
  \bibinfo{author}{\bibfnamefont{V.}~\bibnamefont{Duvvuri}},
  \bibinfo{author}{\bibfnamefont{M.}~\bibnamefont{Trodden}}, \bibnamefont{and}
  \bibinfo{author}{\bibfnamefont{M.~S.} \bibnamefont{Turner}},
  \bibinfo{journal}{Phys. Rev.} \textbf{\bibinfo{volume}{D70}},
  \bibinfo{pages}{043528} (\bibinfo{year}{2004}), \eprint{astro-ph/0306438}.

\bibitem[{\citenamefont{Nojiri and Odintsov}(2007)}]{Nojiri:2006ri}
\bibinfo{author}{\bibfnamefont{S.}~\bibnamefont{Nojiri}} \bibnamefont{and}
  \bibinfo{author}{\bibfnamefont{S.~D.} \bibnamefont{Odintsov}},
  \bibinfo{journal}{Int. J. Geom. Meth. Mod. Phys.}
  \textbf{\bibinfo{volume}{4}}, \bibinfo{pages}{115} (\bibinfo{year}{2007}),
  \eprint{hep-th/0601213}.

\bibitem[{\citenamefont{Capozziello}(2002)}]{Capozziello:2002rd}
\bibinfo{author}{\bibfnamefont{S.}~\bibnamefont{Capozziello}},
  \bibinfo{journal}{Int. J. Mod. Phys.} \textbf{\bibinfo{volume}{D11}},
  \bibinfo{pages}{483} (\bibinfo{year}{2002}), \eprint{gr-qc/0201033}.

\bibitem[{\citenamefont{Capozziello et~al.}(2003)\citenamefont{Capozziello,
  Carloni, and Troisi}}]{Capozziello:2003tk}
\bibinfo{author}{\bibfnamefont{S.}~\bibnamefont{Capozziello}},
  \bibinfo{author}{\bibfnamefont{S.}~\bibnamefont{Carloni}}, \bibnamefont{and}
  \bibinfo{author}{\bibfnamefont{A.}~\bibnamefont{Troisi}}
  (\bibinfo{year}{2003}), \eprint{astro-ph/0303041}.

\bibitem[{\citenamefont{Nojiri and
  Odintsov}(2003{\natexlab{a}})}]{Nojiri:2003ft}
\bibinfo{author}{\bibfnamefont{S.}~\bibnamefont{Nojiri}} \bibnamefont{and}
  \bibinfo{author}{\bibfnamefont{S.~D.} \bibnamefont{Odintsov}},
  \bibinfo{journal}{Phys. Rev.} \textbf{\bibinfo{volume}{D68}},
  \bibinfo{pages}{123512} (\bibinfo{year}{2003}{\natexlab{a}}),
  \eprint{hep-th/0307288}.

\bibitem[{\citenamefont{Nojiri and
  Odintsov}(2003{\natexlab{b}})}]{Nojiri:2003rz}
\bibinfo{author}{\bibfnamefont{S.}~\bibnamefont{Nojiri}} \bibnamefont{and}
  \bibinfo{author}{\bibfnamefont{S.~D.} \bibnamefont{Odintsov}},
  \bibinfo{journal}{Phys. Lett.} \textbf{\bibinfo{volume}{B576}},
  \bibinfo{pages}{5} (\bibinfo{year}{2003}{\natexlab{b}}),
  \eprint{hep-th/0307071}.

\bibitem[{\citenamefont{Faraoni}(2005)}]{Faraoni:2005vk}
\bibinfo{author}{\bibfnamefont{V.}~\bibnamefont{Faraoni}},
  \bibinfo{journal}{Phys. Rev.} \textbf{\bibinfo{volume}{D72}},
  \bibinfo{pages}{124005} (\bibinfo{year}{2005}), \eprint{gr-qc/0511094}.

\bibitem[{\citenamefont{de~la Cruz-Dombriz and
  Dobado}(2006)}]{delaCruz-Dombriz:2006fj}
\bibinfo{author}{\bibfnamefont{A.}~\bibnamefont{de~la Cruz-Dombriz}}
  \bibnamefont{and} \bibinfo{author}{\bibfnamefont{A.}~\bibnamefont{Dobado}},
  \bibinfo{journal}{Phys. Rev.} \textbf{\bibinfo{volume}{D74}},
  \bibinfo{pages}{087501} (\bibinfo{year}{2006}), \eprint{gr-qc/0607118}.

\bibitem[{\citenamefont{Poplawski}(2006)}]{Poplawski:2006kv}
\bibinfo{author}{\bibfnamefont{N.~J.} \bibnamefont{Poplawski}},
  \bibinfo{journal}{Phys. Rev.} \textbf{\bibinfo{volume}{D74}},
  \bibinfo{pages}{084032} (\bibinfo{year}{2006}), \eprint{gr-qc/0607124}.

\bibitem[{\citenamefont{Brookfield et~al.}(2006)\citenamefont{Brookfield,
  van~de Bruck, and Hall}}]{Brookfield:2006mq}
\bibinfo{author}{\bibfnamefont{A.~W.} \bibnamefont{Brookfield}},
  \bibinfo{author}{\bibfnamefont{C.}~\bibnamefont{van~de Bruck}},
  \bibnamefont{and} \bibinfo{author}{\bibfnamefont{L.~M.~H.}
  \bibnamefont{Hall}}, \bibinfo{journal}{Phys. Rev.}
  \textbf{\bibinfo{volume}{D74}}, \bibinfo{pages}{064028}
  (\bibinfo{year}{2006}), \eprint{hep-th/0608015}.

\bibitem[{\citenamefont{Li et~al.}(2006)\citenamefont{Li, Chan, and
  Chu}}]{Li:2006ag}
\bibinfo{author}{\bibfnamefont{B.}~\bibnamefont{Li}},
  \bibinfo{author}{\bibfnamefont{K.~C.} \bibnamefont{Chan}}, \bibnamefont{and}
  \bibinfo{author}{\bibfnamefont{M.~C.} \bibnamefont{Chu}}
  (\bibinfo{year}{2006}), \eprint{astro-ph/0610794}.

\bibitem[{\citenamefont{Sotiriou and Liberati}(2007)}]{Sotiriou:2006qn}
\bibinfo{author}{\bibfnamefont{T.~P.} \bibnamefont{Sotiriou}} \bibnamefont{and}
  \bibinfo{author}{\bibfnamefont{S.}~\bibnamefont{Liberati}},
  \bibinfo{journal}{Annals Phys.} \textbf{\bibinfo{volume}{322}},
  \bibinfo{pages}{935} (\bibinfo{year}{2007}), \eprint{gr-qc/0604006}.

\bibitem[{\citenamefont{Sotiriou}(2007)}]{Sotiriou:2006sf}
\bibinfo{author}{\bibfnamefont{T.~P.} \bibnamefont{Sotiriou}},
  \bibinfo{journal}{Phys. Lett.} \textbf{\bibinfo{volume}{B645}},
  \bibinfo{pages}{389} (\bibinfo{year}{2007}), \eprint{gr-qc/0611107}.

\bibitem[{\citenamefont{Sotiriou}(2006{\natexlab{a}})}]{Sotiriou:2006hs}
\bibinfo{author}{\bibfnamefont{T.~P.} \bibnamefont{Sotiriou}},
  \bibinfo{journal}{Class. Quant. Grav.} \textbf{\bibinfo{volume}{23}},
  \bibinfo{pages}{5117} (\bibinfo{year}{2006}{\natexlab{a}}),
  \eprint{gr-qc/0604028}.

\bibitem[{\citenamefont{Bean et~al.}(2007)\citenamefont{Bean, Bernat, Pogosian,
  Silvestri, and Trodden}}]{Bean:2006up}
\bibinfo{author}{\bibfnamefont{R.}~\bibnamefont{Bean}},
  \bibinfo{author}{\bibfnamefont{D.}~\bibnamefont{Bernat}},
  \bibinfo{author}{\bibfnamefont{L.}~\bibnamefont{Pogosian}},
  \bibinfo{author}{\bibfnamefont{A.}~\bibnamefont{Silvestri}},
  \bibnamefont{and} \bibinfo{author}{\bibfnamefont{M.}~\bibnamefont{Trodden}},
  \bibinfo{journal}{Phys. Rev.} \textbf{\bibinfo{volume}{D75}},
  \bibinfo{pages}{064020} (\bibinfo{year}{2007}), \eprint{astro-ph/0611321}.

\bibitem[{\citenamefont{Baghram et~al.}(2007)\citenamefont{Baghram, Farhang,
  and Rahvar}}]{Baghram:2007df}
\bibinfo{author}{\bibfnamefont{S.}~\bibnamefont{Baghram}},
  \bibinfo{author}{\bibfnamefont{M.}~\bibnamefont{Farhang}}, \bibnamefont{and}
  \bibinfo{author}{\bibfnamefont{S.}~\bibnamefont{Rahvar}},
  \bibinfo{journal}{Phys. Rev.} \textbf{\bibinfo{volume}{D75}},
  \bibinfo{pages}{044024} (\bibinfo{year}{2007}), \eprint{astro-ph/0701013}.

\bibitem[{\citenamefont{Bazeia et~al.}(2007)\citenamefont{Bazeia, Carneiro~da
  Cunha, Menezes, and Petrov}}]{Bazeia:2007jj}
\bibinfo{author}{\bibfnamefont{D.}~\bibnamefont{Bazeia}},
  \bibinfo{author}{\bibfnamefont{B.}~\bibnamefont{Carneiro~da Cunha}},
  \bibinfo{author}{\bibfnamefont{R.}~\bibnamefont{Menezes}}, \bibnamefont{and}
  \bibinfo{author}{\bibfnamefont{A.~Y.} \bibnamefont{Petrov}}
  (\bibinfo{year}{2007}), \eprint{hep-th/0701106}.

\bibitem[{\citenamefont{Li and Barrow}(2007)}]{Li:2007xn}
\bibinfo{author}{\bibfnamefont{B.}~\bibnamefont{Li}} \bibnamefont{and}
  \bibinfo{author}{\bibfnamefont{J.~D.} \bibnamefont{Barrow}},
  \bibinfo{journal}{Phys. Rev.} \textbf{\bibinfo{volume}{D75}},
  \bibinfo{pages}{084010} (\bibinfo{year}{2007}), \eprint{gr-qc/0701111}.

\bibitem[{\citenamefont{Bludman}(2007)}]{Bludman:2007kg}
\bibinfo{author}{\bibfnamefont{S.}~\bibnamefont{Bludman}}
  (\bibinfo{year}{2007}), \eprint{astro-ph/0702085}.

\bibitem[{\citenamefont{Rador}(2007{\natexlab{a}})}]{Rador:2007wq}
\bibinfo{author}{\bibfnamefont{T.}~\bibnamefont{Rador}}
  (\bibinfo{year}{2007}{\natexlab{a}}), \eprint{hep-th/0702081}.

\bibitem[{\citenamefont{Rador}(2007{\natexlab{b}})}]{Rador:2007gq}
\bibinfo{author}{\bibfnamefont{T.}~\bibnamefont{Rador}},
  \bibinfo{journal}{Phys. Rev.} \textbf{\bibinfo{volume}{D75}},
  \bibinfo{pages}{064033} (\bibinfo{year}{2007}{\natexlab{b}}),
  \eprint{hep-th/0701267}.

\bibitem[{\citenamefont{Sokolowski}(2007)}]{Sokolowski:2007pk}
\bibinfo{author}{\bibfnamefont{L.~M.} \bibnamefont{Sokolowski}}
  (\bibinfo{year}{2007}), \eprint{gr-qc/0702097}.

\bibitem[{\citenamefont{Faraoni}(2007)}]{Faraoni:2007yn}
\bibinfo{author}{\bibfnamefont{V.}~\bibnamefont{Faraoni}},
  \bibinfo{journal}{Phys. Rev.} \textbf{\bibinfo{volume}{D75}},
  \bibinfo{pages}{067302} (\bibinfo{year}{2007}), \eprint{gr-qc/0703044}.

\bibitem[{\citenamefont{Faraoni}(2006)}]{Faraoni:2006sy}
\bibinfo{author}{\bibfnamefont{V.}~\bibnamefont{Faraoni}},
  \bibinfo{journal}{Phys. Rev.} \textbf{\bibinfo{volume}{D74}},
  \bibinfo{pages}{104017} (\bibinfo{year}{2006}), \eprint{astro-ph/0610734}.

\bibitem[{\citenamefont{Nojiri and Odintsov}(2004)}]{Nojiri:2003ni}
\bibinfo{author}{\bibfnamefont{S.}~\bibnamefont{Nojiri}} \bibnamefont{and}
  \bibinfo{author}{\bibfnamefont{S.~D.} \bibnamefont{Odintsov}},
  \bibinfo{journal}{Gen. Rel. Grav.} \textbf{\bibinfo{volume}{36}},
  \bibinfo{pages}{1765} (\bibinfo{year}{2004}), \eprint{hep-th/0308176}.

\bibitem[{\citenamefont{Wang and Meng}(2004)}]{Wang:2004vs}
\bibinfo{author}{\bibfnamefont{P.}~\bibnamefont{Wang}} \bibnamefont{and}
  \bibinfo{author}{\bibfnamefont{X.-H.} \bibnamefont{Meng}},
  \bibinfo{journal}{TSPU Vestnik} \textbf{\bibinfo{volume}{44N7}},
  \bibinfo{pages}{40} (\bibinfo{year}{2004}), \eprint{astro-ph/0406455}.

\bibitem[{\citenamefont{Meng and Wang}(2004)}]{Meng:2003sx}
\bibinfo{author}{\bibfnamefont{X.-H.} \bibnamefont{Meng}} \bibnamefont{and}
  \bibinfo{author}{\bibfnamefont{P.}~\bibnamefont{Wang}},
  \bibinfo{journal}{Gen. Rel. Grav.} \textbf{\bibinfo{volume}{36}},
  \bibinfo{pages}{1947} (\bibinfo{year}{2004}), \eprint{gr-qc/0311019}.

\bibitem[{\citenamefont{Abdalla et~al.}(2005)\citenamefont{Abdalla, Nojiri, and
  Odintsov}}]{Abdalla:2004sw}
\bibinfo{author}{\bibfnamefont{M.~C.~B.} \bibnamefont{Abdalla}},
  \bibinfo{author}{\bibfnamefont{S.}~\bibnamefont{Nojiri}}, \bibnamefont{and}
  \bibinfo{author}{\bibfnamefont{S.~D.} \bibnamefont{Odintsov}},
  \bibinfo{journal}{Class. Quant. Grav.} \textbf{\bibinfo{volume}{22}},
  \bibinfo{pages}{L35} (\bibinfo{year}{2005}), \eprint{hep-th/0409177}.

\bibitem[{\citenamefont{Cognola et~al.}(2005)\citenamefont{Cognola, Elizalde,
  Nojiri, Odintsov, and Zerbini}}]{Cognola:2005de}
\bibinfo{author}{\bibfnamefont{G.}~\bibnamefont{Cognola}},
  \bibinfo{author}{\bibfnamefont{E.}~\bibnamefont{Elizalde}},
  \bibinfo{author}{\bibfnamefont{S.}~\bibnamefont{Nojiri}},
  \bibinfo{author}{\bibfnamefont{S.~D.} \bibnamefont{Odintsov}},
  \bibnamefont{and} \bibinfo{author}{\bibfnamefont{S.}~\bibnamefont{Zerbini}},
  \bibinfo{journal}{JCAP} \textbf{\bibinfo{volume}{0502}}, \bibinfo{pages}{010}
  (\bibinfo{year}{2005}), \eprint{hep-th/0501096}.

\bibitem[{\citenamefont{Capozziello et~al.}(2005)\citenamefont{Capozziello,
  Cardone, and Troisi}}]{Capozziello:2005ku}
\bibinfo{author}{\bibfnamefont{S.}~\bibnamefont{Capozziello}},
  \bibinfo{author}{\bibfnamefont{V.~F.} \bibnamefont{Cardone}},
  \bibnamefont{and} \bibinfo{author}{\bibfnamefont{A.}~\bibnamefont{Troisi}},
  \bibinfo{journal}{Phys. Rev.} \textbf{\bibinfo{volume}{D71}},
  \bibinfo{pages}{043503} (\bibinfo{year}{2005}), \eprint{astro-ph/0501426}.

\bibitem[{\citenamefont{Allemandi
  et~al.}(2005{\natexlab{a}})\citenamefont{Allemandi, Borowiec, Francaviglia,
  and Odintsov}}]{Allemandi:2005qs}
\bibinfo{author}{\bibfnamefont{G.}~\bibnamefont{Allemandi}},
  \bibinfo{author}{\bibfnamefont{A.}~\bibnamefont{Borowiec}},
  \bibinfo{author}{\bibfnamefont{M.}~\bibnamefont{Francaviglia}},
  \bibnamefont{and} \bibinfo{author}{\bibfnamefont{S.~D.}
  \bibnamefont{Odintsov}}, \bibinfo{journal}{Phys. Rev.}
  \textbf{\bibinfo{volume}{D72}}, \bibinfo{pages}{063505}
  (\bibinfo{year}{2005}{\natexlab{a}}), \eprint{gr-qc/0504057}.

\bibitem[{\citenamefont{Koivisto and Kurki-Suonio}(2006)}]{Koivisto:2005yc}
\bibinfo{author}{\bibfnamefont{T.}~\bibnamefont{Koivisto}} \bibnamefont{and}
  \bibinfo{author}{\bibfnamefont{H.}~\bibnamefont{Kurki-Suonio}},
  \bibinfo{journal}{Class. Quant. Grav.} \textbf{\bibinfo{volume}{23}},
  \bibinfo{pages}{2355} (\bibinfo{year}{2006}), \eprint{astro-ph/0509422}.

\bibitem[{\citenamefont{Clifton and Barrow}(2005)}]{Clifton:2005aj}
\bibinfo{author}{\bibfnamefont{T.}~\bibnamefont{Clifton}} \bibnamefont{and}
  \bibinfo{author}{\bibfnamefont{J.~D.} \bibnamefont{Barrow}},
  \bibinfo{journal}{Phys. Rev.} \textbf{\bibinfo{volume}{D72}},
  \bibinfo{pages}{103005} (\bibinfo{year}{2005}), \eprint{gr-qc/0509059}.

\bibitem[{\citenamefont{Mena et~al.}(2006)\citenamefont{Mena, Santiago, and
  Weller}}]{Mena:2005ta}
\bibinfo{author}{\bibfnamefont{O.}~\bibnamefont{Mena}},
  \bibinfo{author}{\bibfnamefont{J.}~\bibnamefont{Santiago}}, \bibnamefont{and}
  \bibinfo{author}{\bibfnamefont{J.}~\bibnamefont{Weller}},
  \bibinfo{journal}{Phys. Rev. Lett.} \textbf{\bibinfo{volume}{96}},
  \bibinfo{pages}{041103} (\bibinfo{year}{2006}), \eprint{astro-ph/0510453}.

\bibitem[{\citenamefont{Amarzguioui et~al.}(2006)\citenamefont{Amarzguioui,
  Elgaroy, Mota, and Multamaki}}]{Amarzguioui:2005zq}
\bibinfo{author}{\bibfnamefont{M.}~\bibnamefont{Amarzguioui}},
  \bibinfo{author}{\bibfnamefont{O.}~\bibnamefont{Elgaroy}},
  \bibinfo{author}{\bibfnamefont{D.~F.} \bibnamefont{Mota}}, \bibnamefont{and}
  \bibinfo{author}{\bibfnamefont{T.}~\bibnamefont{Multamaki}},
  \bibinfo{journal}{Astron. Astrophys.} \textbf{\bibinfo{volume}{454}},
  \bibinfo{pages}{707} (\bibinfo{year}{2006}), \eprint{astro-ph/0510519}.

\bibitem[{\citenamefont{Brevik}(2006)}]{Brevik:2006md}
\bibinfo{author}{\bibfnamefont{I.}~\bibnamefont{Brevik}},
  \bibinfo{journal}{Int. J. Mod. Phys.} \textbf{\bibinfo{volume}{D15}},
  \bibinfo{pages}{767} (\bibinfo{year}{2006}), \eprint{gr-qc/0601100}.

\bibitem[{\citenamefont{Koivisto}(2006)}]{Koivisto:2006ie}
\bibinfo{author}{\bibfnamefont{T.}~\bibnamefont{Koivisto}},
  \bibinfo{journal}{Phys. Rev.} \textbf{\bibinfo{volume}{D73}},
  \bibinfo{pages}{083517} (\bibinfo{year}{2006}), \eprint{astro-ph/0602031}.

\bibitem[{\citenamefont{Perez~Bergliaffa}(2006)}]{PerezBergliaffa:2006ni}
\bibinfo{author}{\bibfnamefont{S.~E.} \bibnamefont{Perez~Bergliaffa}},
  \bibinfo{journal}{Phys. Lett.} \textbf{\bibinfo{volume}{B642}},
  \bibinfo{pages}{311} (\bibinfo{year}{2006}), \eprint{gr-qc/0608072}.

\bibitem[{\citenamefont{Cognola
  et~al.}(2007{\natexlab{a}})\citenamefont{Cognola, Gastaldi, and
  Zerbini}}]{Cognola:2007vq}
\bibinfo{author}{\bibfnamefont{G.}~\bibnamefont{Cognola}},
  \bibinfo{author}{\bibfnamefont{M.}~\bibnamefont{Gastaldi}}, \bibnamefont{and}
  \bibinfo{author}{\bibfnamefont{S.}~\bibnamefont{Zerbini}}
  (\bibinfo{year}{2007}{\natexlab{a}}), \eprint{gr-qc/0701138}.

\bibitem[{\citenamefont{Capozziello and Garattini}(2007)}]{Capozziello:2007gm}
\bibinfo{author}{\bibfnamefont{S.}~\bibnamefont{Capozziello}} \bibnamefont{and}
  \bibinfo{author}{\bibfnamefont{R.}~\bibnamefont{Garattini}},
  \bibinfo{journal}{Class. Quant. Grav.} \textbf{\bibinfo{volume}{24}},
  \bibinfo{pages}{1627} (\bibinfo{year}{2007}), \eprint{gr-qc/0702075}.

\bibitem[{\citenamefont{Nojiri and
  Odintsov}(2006{\natexlab{a}})}]{Nojiri:2006gh}
\bibinfo{author}{\bibfnamefont{S.}~\bibnamefont{Nojiri}} \bibnamefont{and}
  \bibinfo{author}{\bibfnamefont{S.~D.} \bibnamefont{Odintsov}},
  \bibinfo{journal}{Phys. Rev.} \textbf{\bibinfo{volume}{D74}},
  \bibinfo{pages}{086005} (\bibinfo{year}{2006}{\natexlab{a}}),
  \eprint{hep-th/0608008}.

\bibitem[{\citenamefont{Nojiri and
  Odintsov}(2006{\natexlab{b}})}]{Nojiri:2006su}
\bibinfo{author}{\bibfnamefont{S.}~\bibnamefont{Nojiri}} \bibnamefont{and}
  \bibinfo{author}{\bibfnamefont{S.~D.} \bibnamefont{Odintsov}}
  (\bibinfo{year}{2006}{\natexlab{b}}), \eprint{hep-th/0610164}.

\bibitem[{\citenamefont{Capozziello et~al.}(2006)\citenamefont{Capozziello,
  Nojiri, Odintsov, and Troisi}}]{Capozziello:2006dj}
\bibinfo{author}{\bibfnamefont{S.}~\bibnamefont{Capozziello}},
  \bibinfo{author}{\bibfnamefont{S.}~\bibnamefont{Nojiri}},
  \bibinfo{author}{\bibfnamefont{S.~D.} \bibnamefont{Odintsov}},
  \bibnamefont{and} \bibinfo{author}{\bibfnamefont{A.}~\bibnamefont{Troisi}},
  \bibinfo{journal}{Phys. Lett.} \textbf{\bibinfo{volume}{B639}},
  \bibinfo{pages}{135} (\bibinfo{year}{2006}), \eprint{astro-ph/0604431}.

\bibitem[{\citenamefont{Fay et~al.}(2007{\natexlab{a}})\citenamefont{Fay,
  Nesseris, and Perivolaropoulos}}]{Fay:2007uy}
\bibinfo{author}{\bibfnamefont{S.}~\bibnamefont{Fay}},
  \bibinfo{author}{\bibfnamefont{S.}~\bibnamefont{Nesseris}}, \bibnamefont{and}
  \bibinfo{author}{\bibfnamefont{L.}~\bibnamefont{Perivolaropoulos}}
  (\bibinfo{year}{2007}{\natexlab{a}}), \eprint{gr-qc/0703006}.

\bibitem[{\citenamefont{Fay et~al.}(2007{\natexlab{b}})\citenamefont{Fay,
  Tavakol, and Tsujikawa}}]{Fay:2007gg}
\bibinfo{author}{\bibfnamefont{S.}~\bibnamefont{Fay}},
  \bibinfo{author}{\bibfnamefont{R.}~\bibnamefont{Tavakol}}, \bibnamefont{and}
  \bibinfo{author}{\bibfnamefont{S.}~\bibnamefont{Tsujikawa}},
  \bibinfo{journal}{Phys. Rev.} \textbf{\bibinfo{volume}{D75}},
  \bibinfo{pages}{063509} (\bibinfo{year}{2007}{\natexlab{b}}),
  \eprint{astro-ph/0701479}.

\bibitem[{\citenamefont{Nojiri et~al.}(2005)\citenamefont{Nojiri, Odintsov, and
  Sasaki}}]{Nojiri:2005vv}
\bibinfo{author}{\bibfnamefont{S.}~\bibnamefont{Nojiri}},
  \bibinfo{author}{\bibfnamefont{S.~D.} \bibnamefont{Odintsov}},
  \bibnamefont{and} \bibinfo{author}{\bibfnamefont{M.}~\bibnamefont{Sasaki}},
  \bibinfo{journal}{Phys. Rev.} \textbf{\bibinfo{volume}{D71}},
  \bibinfo{pages}{123509} (\bibinfo{year}{2005}), \eprint{hep-th/0504052}.

\bibitem[{\citenamefont{Sami et~al.}(2005)\citenamefont{Sami, Toporensky,
  Tretjakov, and Tsujikawa}}]{Sami:2005zc}
\bibinfo{author}{\bibfnamefont{M.}~\bibnamefont{Sami}},
  \bibinfo{author}{\bibfnamefont{A.}~\bibnamefont{Toporensky}},
  \bibinfo{author}{\bibfnamefont{P.~V.} \bibnamefont{Tretjakov}},
  \bibnamefont{and}
  \bibinfo{author}{\bibfnamefont{S.}~\bibnamefont{Tsujikawa}},
  \bibinfo{journal}{Phys. Lett.} \textbf{\bibinfo{volume}{B619}},
  \bibinfo{pages}{193} (\bibinfo{year}{2005}), \eprint{hep-th/0504154}.

\bibitem[{\citenamefont{Calcagni et~al.}(2005)\citenamefont{Calcagni,
  Tsujikawa, and Sami}}]{Calcagni:2005im}
\bibinfo{author}{\bibfnamefont{G.}~\bibnamefont{Calcagni}},
  \bibinfo{author}{\bibfnamefont{S.}~\bibnamefont{Tsujikawa}},
  \bibnamefont{and} \bibinfo{author}{\bibfnamefont{M.}~\bibnamefont{Sami}},
  \bibinfo{journal}{Class. Quant. Grav.} \textbf{\bibinfo{volume}{22}},
  \bibinfo{pages}{3977} (\bibinfo{year}{2005}), \eprint{hep-th/0505193}.

\bibitem[{\citenamefont{Tsujikawa and Sami}(2007)}]{Tsujikawa:2006ph}
\bibinfo{author}{\bibfnamefont{S.}~\bibnamefont{Tsujikawa}} \bibnamefont{and}
  \bibinfo{author}{\bibfnamefont{M.}~\bibnamefont{Sami}},
  \bibinfo{journal}{JCAP} \textbf{\bibinfo{volume}{0701}}, \bibinfo{pages}{006}
  (\bibinfo{year}{2007}), \eprint{hep-th/0608178}.

\bibitem[{\citenamefont{Guo et~al.}(2007)\citenamefont{Guo, Ohta, and
  Tsujikawa}}]{Guo:2006ct}
\bibinfo{author}{\bibfnamefont{Z.-K.} \bibnamefont{Guo}},
  \bibinfo{author}{\bibfnamefont{N.}~\bibnamefont{Ohta}}, \bibnamefont{and}
  \bibinfo{author}{\bibfnamefont{S.}~\bibnamefont{Tsujikawa}},
  \bibinfo{journal}{Phys. Rev.} \textbf{\bibinfo{volume}{D75}},
  \bibinfo{pages}{023520} (\bibinfo{year}{2007}), \eprint{hep-th/0610336}.

\bibitem[{\citenamefont{Sanyal}(2007)}]{Sanyal:2006wi}
\bibinfo{author}{\bibfnamefont{A.~K.} \bibnamefont{Sanyal}},
  \bibinfo{journal}{Phys. Lett.} \textbf{\bibinfo{volume}{B645}},
  \bibinfo{pages}{1} (\bibinfo{year}{2007}), \eprint{astro-ph/0608104}.

\bibitem[{\citenamefont{Leith and Neupane}(2007)}]{Leith:2007bu}
\bibinfo{author}{\bibfnamefont{B.~M.} \bibnamefont{Leith}} \bibnamefont{and}
  \bibinfo{author}{\bibfnamefont{I.~P.} \bibnamefont{Neupane}}
  (\bibinfo{year}{2007}), \eprint{hep-th/0702002}.

\bibitem[{\citenamefont{Carter and Neupane}(2006)}]{Carter:2005fu}
\bibinfo{author}{\bibfnamefont{B.~M.~N.} \bibnamefont{Carter}}
  \bibnamefont{and} \bibinfo{author}{\bibfnamefont{I.~P.}
  \bibnamefont{Neupane}}, \bibinfo{journal}{JCAP}
  \textbf{\bibinfo{volume}{0606}}, \bibinfo{pages}{004} (\bibinfo{year}{2006}),
  \eprint{hep-th/0512262}.

\bibitem[{\citenamefont{Koivisto and
  Mota}(2007{\natexlab{a}})}]{Koivisto:2006ai}
\bibinfo{author}{\bibfnamefont{T.}~\bibnamefont{Koivisto}} \bibnamefont{and}
  \bibinfo{author}{\bibfnamefont{D.~F.} \bibnamefont{Mota}},
  \bibinfo{journal}{Phys. Rev.} \textbf{\bibinfo{volume}{D75}},
  \bibinfo{pages}{023518} (\bibinfo{year}{2007}{\natexlab{a}}),
  \eprint{hep-th/0609155}.

\bibitem[{\citenamefont{Koivisto and
  Mota}(2007{\natexlab{b}})}]{Koivisto:2006xf}
\bibinfo{author}{\bibfnamefont{T.}~\bibnamefont{Koivisto}} \bibnamefont{and}
  \bibinfo{author}{\bibfnamefont{D.~F.} \bibnamefont{Mota}},
  \bibinfo{journal}{Phys. Lett.} \textbf{\bibinfo{volume}{B644}},
  \bibinfo{pages}{104} (\bibinfo{year}{2007}{\natexlab{b}}),
  \eprint{astro-ph/0606078}.

\bibitem[{\citenamefont{Nojiri et~al.}(2006)\citenamefont{Nojiri, Odintsov, and
  Sami}}]{Nojiri:2006je}
\bibinfo{author}{\bibfnamefont{S.}~\bibnamefont{Nojiri}},
  \bibinfo{author}{\bibfnamefont{S.~D.} \bibnamefont{Odintsov}},
  \bibnamefont{and} \bibinfo{author}{\bibfnamefont{M.}~\bibnamefont{Sami}},
  \bibinfo{journal}{Phys. Rev.} \textbf{\bibinfo{volume}{D74}},
  \bibinfo{pages}{046004} (\bibinfo{year}{2006}), \eprint{hep-th/0605039}.

\bibitem[{\citenamefont{Nojiri and
  Odintsov}(2006{\natexlab{c}})}]{Nojiri:2006be}
\bibinfo{author}{\bibfnamefont{S.}~\bibnamefont{Nojiri}} \bibnamefont{and}
  \bibinfo{author}{\bibfnamefont{S.~D.} \bibnamefont{Odintsov}}
  (\bibinfo{year}{2006}{\natexlab{c}}), \eprint{hep-th/0611071}.

\bibitem[{\citenamefont{Cognola
  et~al.}(2007{\natexlab{b}})\citenamefont{Cognola, Elizalde, Nojiri, Odintsov,
  and Zerbini}}]{Cognola:2006sp}
\bibinfo{author}{\bibfnamefont{G.}~\bibnamefont{Cognola}},
  \bibinfo{author}{\bibfnamefont{E.}~\bibnamefont{Elizalde}},
  \bibinfo{author}{\bibfnamefont{S.}~\bibnamefont{Nojiri}},
  \bibinfo{author}{\bibfnamefont{S.}~\bibnamefont{Odintsov}}, \bibnamefont{and}
  \bibinfo{author}{\bibfnamefont{S.}~\bibnamefont{Zerbini}},
  \bibinfo{journal}{Phys. Rev.} \textbf{\bibinfo{volume}{D75}},
  \bibinfo{pages}{086002} (\bibinfo{year}{2007}{\natexlab{b}}),
  \eprint{hep-th/0611198}.

\bibitem[{\citenamefont{Nojiri and Odintsov}(2005)}]{Nojiri:2005jg}
\bibinfo{author}{\bibfnamefont{S.}~\bibnamefont{Nojiri}} \bibnamefont{and}
  \bibinfo{author}{\bibfnamefont{S.~D.} \bibnamefont{Odintsov}},
  \bibinfo{journal}{Phys. Lett.} \textbf{\bibinfo{volume}{B631}},
  \bibinfo{pages}{1} (\bibinfo{year}{2005}), \eprint{hep-th/0508049}.

\bibitem[{\citenamefont{Cognola et~al.}(2006)\citenamefont{Cognola, Elizalde,
  Nojiri, Odintsov, and Zerbini}}]{Cognola:2006eg}
\bibinfo{author}{\bibfnamefont{G.}~\bibnamefont{Cognola}},
  \bibinfo{author}{\bibfnamefont{E.}~\bibnamefont{Elizalde}},
  \bibinfo{author}{\bibfnamefont{S.}~\bibnamefont{Nojiri}},
  \bibinfo{author}{\bibfnamefont{S.~D.} \bibnamefont{Odintsov}},
  \bibnamefont{and} \bibinfo{author}{\bibfnamefont{S.}~\bibnamefont{Zerbini}},
  \bibinfo{journal}{Phys. Rev.} \textbf{\bibinfo{volume}{D73}},
  \bibinfo{pages}{084007} (\bibinfo{year}{2006}), \eprint{hep-th/0601008}.

\bibitem[{\citenamefont{Song et~al.}(2007)\citenamefont{Song, Hu, and
  Sawicki}}]{SonHuSaw06}
\bibinfo{author}{\bibfnamefont{Y.-S.} \bibnamefont{Song}},
  \bibinfo{author}{\bibfnamefont{W.}~\bibnamefont{Hu}}, \bibnamefont{and}
  \bibinfo{author}{\bibfnamefont{I.}~\bibnamefont{Sawicki}},
  \bibinfo{journal}{Phys. Rev.} \textbf{\bibinfo{volume}{D75}},
  \bibinfo{pages}{044004} (\bibinfo{year}{2007}), \eprint{astro-ph/0610532}.

\bibitem[{\citenamefont{Nojiri et~al.}(2007)\citenamefont{Nojiri, Odintsov, and
  Tretyakov}}]{Nojiri:2007te}
\bibinfo{author}{\bibfnamefont{S.}~\bibnamefont{Nojiri}},
  \bibinfo{author}{\bibfnamefont{S.~D.} \bibnamefont{Odintsov}},
  \bibnamefont{and} \bibinfo{author}{\bibfnamefont{P.~V.}
  \bibnamefont{Tretyakov}} (\bibinfo{year}{2007}), \eprint{arXiv:0704.2520
  [hep-th]}.

\bibitem[{\citenamefont{Multamaki and Vilja}(2006)}]{MulVil06a}
\bibinfo{author}{\bibfnamefont{T.}~\bibnamefont{Multamaki}} \bibnamefont{and}
  \bibinfo{author}{\bibfnamefont{I.}~\bibnamefont{Vilja}},
  \bibinfo{journal}{Phys. Rev.} \textbf{\bibinfo{volume}{D73}},
  \bibinfo{pages}{024018} (\bibinfo{year}{2006}), \eprint{astro-ph/0506692}.

\bibitem[{\citenamefont{Woodard}(2006)}]{Woo06}
\bibinfo{author}{\bibfnamefont{R.~P.} \bibnamefont{Woodard}}
  (\bibinfo{year}{2006}), \eprint{astro-ph/0601672}.

\bibitem[{\citenamefont{Sotiriou}(2006{\natexlab{b}})}]{Sot05}
\bibinfo{author}{\bibfnamefont{T.~P.} \bibnamefont{Sotiriou}},
  \bibinfo{journal}{Gen. Rel. Grav.} \textbf{\bibinfo{volume}{38}},
  \bibinfo{pages}{1407} (\bibinfo{year}{2006}{\natexlab{b}}),
  \eprint{gr-qc/0507027}.

\bibitem[{\citenamefont{Cembranos}(2006)}]{Cem05}
\bibinfo{author}{\bibfnamefont{J.~A.~R.} \bibnamefont{Cembranos}},
  \bibinfo{journal}{Phys. Rev.} \textbf{\bibinfo{volume}{D73}},
  \bibinfo{pages}{064029} (\bibinfo{year}{2006}), \eprint{gr-qc/0507039}.

\bibitem[{\citenamefont{Allemandi
  et~al.}(2005{\natexlab{b}})\citenamefont{Allemandi, Francaviglia, Ruggiero,
  and Tartaglia}}]{Allemandi:2005tg}
\bibinfo{author}{\bibfnamefont{G.}~\bibnamefont{Allemandi}},
  \bibinfo{author}{\bibfnamefont{M.}~\bibnamefont{Francaviglia}},
  \bibinfo{author}{\bibfnamefont{M.~L.} \bibnamefont{Ruggiero}},
  \bibnamefont{and}
  \bibinfo{author}{\bibfnamefont{A.}~\bibnamefont{Tartaglia}},
  \bibinfo{journal}{Gen. Rel. Grav.} \textbf{\bibinfo{volume}{37}},
  \bibinfo{pages}{1891} (\bibinfo{year}{2005}{\natexlab{b}}),
  \eprint{gr-qc/0506123}.

\bibitem[{\citenamefont{Ruggiero and Iorio}(2007)}]{Ruggiero:2006qv}
\bibinfo{author}{\bibfnamefont{M.~L.} \bibnamefont{Ruggiero}} \bibnamefont{and}
  \bibinfo{author}{\bibfnamefont{L.}~\bibnamefont{Iorio}},
  \bibinfo{journal}{JCAP} \textbf{\bibinfo{volume}{0701}}, \bibinfo{pages}{010}
  (\bibinfo{year}{2007}), \eprint{gr-qc/0607093}.

\bibitem[{\citenamefont{Sotiriou and Barausse}(2007)}]{Sotiriou:2006pq}
\bibinfo{author}{\bibfnamefont{T.~P.} \bibnamefont{Sotiriou}} \bibnamefont{and}
  \bibinfo{author}{\bibfnamefont{E.}~\bibnamefont{Barausse}},
  \bibinfo{journal}{Phys. Rev.} \textbf{\bibinfo{volume}{D75}},
  \bibinfo{pages}{084007} (\bibinfo{year}{2007}), \eprint{gr-qc/0612065}.

\bibitem[{\citenamefont{Shao et~al.}(2006)\citenamefont{Shao, Cai, Wang, and
  Su}}]{ShaCaiWanSu06}
\bibinfo{author}{\bibfnamefont{C.-G.} \bibnamefont{Shao}},
  \bibinfo{author}{\bibfnamefont{R.-G.} \bibnamefont{Cai}},
  \bibinfo{author}{\bibfnamefont{B.}~\bibnamefont{Wang}}, \bibnamefont{and}
  \bibinfo{author}{\bibfnamefont{R.-K.} \bibnamefont{Su}},
  \bibinfo{journal}{Phys. Lett.} \textbf{\bibinfo{volume}{B633}},
  \bibinfo{pages}{164} (\bibinfo{year}{2006}), \eprint{gr-qc/0511034}.

\bibitem[{\citenamefont{Bustelo and Barraco}(2007)}]{Bustelo:2006ms}
\bibinfo{author}{\bibfnamefont{A.~J.} \bibnamefont{Bustelo}} \bibnamefont{and}
  \bibinfo{author}{\bibfnamefont{D.~E.} \bibnamefont{Barraco}},
  \bibinfo{journal}{Class. Quant. Grav.} \textbf{\bibinfo{volume}{24}},
  \bibinfo{pages}{2333} (\bibinfo{year}{2007}), \eprint{gr-qc/0611149}.

\bibitem[{\citenamefont{Olmo}(2007)}]{Olmo:2006eh}
\bibinfo{author}{\bibfnamefont{G.~J.} \bibnamefont{Olmo}},
  \bibinfo{journal}{Phys. Rev.} \textbf{\bibinfo{volume}{D75}},
  \bibinfo{pages}{023511} (\bibinfo{year}{2007}), \eprint{gr-qc/0612047}.

\bibitem[{\citenamefont{Zhang}(2007)}]{Zha07}
\bibinfo{author}{\bibfnamefont{P.-J.} \bibnamefont{Zhang}}
  (\bibinfo{year}{2007}), \eprint{astro-ph/0701662}.

\bibitem[{\citenamefont{Kainulainen et~al.}(2007)\citenamefont{Kainulainen,
  Piilonen, Reijonen, and Sunhede}}]{Kainulainen:2007bt}
\bibinfo{author}{\bibfnamefont{K.}~\bibnamefont{Kainulainen}},
  \bibinfo{author}{\bibfnamefont{J.}~\bibnamefont{Piilonen}},
  \bibinfo{author}{\bibfnamefont{V.}~\bibnamefont{Reijonen}}, \bibnamefont{and}
  \bibinfo{author}{\bibfnamefont{D.}~\bibnamefont{Sunhede}}
  (\bibinfo{year}{2007}), \eprint{arXiv:0704.2729 [gr-qc]}.

\bibitem[{\citenamefont{Chiba}(2003)}]{Chi03}
\bibinfo{author}{\bibfnamefont{T.}~\bibnamefont{Chiba}},
  \bibinfo{journal}{Phys. Lett.} \textbf{\bibinfo{volume}{B575}},
  \bibinfo{pages}{1} (\bibinfo{year}{2003}), \eprint{astro-ph/0307338}.

\bibitem[{\citenamefont{Erickcek et~al.}(2006)\citenamefont{Erickcek, Smith,
  and Kamionkowski}}]{EriSmiKam06}
\bibinfo{author}{\bibfnamefont{A.~L.} \bibnamefont{Erickcek}},
  \bibinfo{author}{\bibfnamefont{T.~L.} \bibnamefont{Smith}}, \bibnamefont{and}
  \bibinfo{author}{\bibfnamefont{M.}~\bibnamefont{Kamionkowski}},
  \bibinfo{journal}{Phys. Rev.} \textbf{\bibinfo{volume}{D74}},
  \bibinfo{pages}{121501} (\bibinfo{year}{2006}), \eprint{astro-ph/0610483}.

\bibitem[{\citenamefont{Chiba et~al.}(2006)\citenamefont{Chiba, Smith, and
  Erickcek}}]{Chiba:2006jp}
\bibinfo{author}{\bibfnamefont{T.}~\bibnamefont{Chiba}},
  \bibinfo{author}{\bibfnamefont{T.~L.} \bibnamefont{Smith}}, \bibnamefont{and}
  \bibinfo{author}{\bibfnamefont{A.~L.} \bibnamefont{Erickcek}}
  (\bibinfo{year}{2006}), \eprint{astro-ph/0611867}.

\bibitem[{\citenamefont{Khoury and Weltman}(2004)}]{KhoWel04}
\bibinfo{author}{\bibfnamefont{J.}~\bibnamefont{Khoury}} \bibnamefont{and}
  \bibinfo{author}{\bibfnamefont{A.}~\bibnamefont{Weltman}},
  \bibinfo{journal}{Phys. Rev.} \textbf{\bibinfo{volume}{D69}},
  \bibinfo{pages}{044026} (\bibinfo{year}{2004}), \eprint{astro-ph/0309411}.

\bibitem[{\citenamefont{Navarro and Van~Acoleyen}(2006)}]{NavAco06}
\bibinfo{author}{\bibfnamefont{I.}~\bibnamefont{Navarro}} \bibnamefont{and}
  \bibinfo{author}{\bibfnamefont{K.}~\bibnamefont{Van~Acoleyen}}
  (\bibinfo{year}{2006}), \eprint{gr-qc/0611127}.

\bibitem[{\citenamefont{Faulkner
  et~al.}(2006{\natexlab{a}})\citenamefont{Faulkner, Tegmark, Bunn, and
  Mao}}]{FauTegBunMao06}
\bibinfo{author}{\bibfnamefont{T.}~\bibnamefont{Faulkner}},
  \bibinfo{author}{\bibfnamefont{M.}~\bibnamefont{Tegmark}},
  \bibinfo{author}{\bibfnamefont{E.~F.} \bibnamefont{Bunn}}, \bibnamefont{and}
  \bibinfo{author}{\bibfnamefont{Y.}~\bibnamefont{Mao}}
  (\bibinfo{year}{2006}{\natexlab{a}}), \eprint{astro-ph/0612569}.

\bibitem[{\citenamefont{Amendola et~al.}(2006)\citenamefont{Amendola, Polarski,
  and Tsujikawa}}]{AmePolTsu06a}
\bibinfo{author}{\bibfnamefont{L.}~\bibnamefont{Amendola}},
  \bibinfo{author}{\bibfnamefont{D.}~\bibnamefont{Polarski}}, \bibnamefont{and}
  \bibinfo{author}{\bibfnamefont{S.}~\bibnamefont{Tsujikawa}}
  (\bibinfo{year}{2006}), \eprint{astro-ph/0603703}.

\bibitem[{\citenamefont{{Sawicki} and {Hu}}(2007)}]{SawHu07}
\bibinfo{author}{\bibfnamefont{I.}~\bibnamefont{{Sawicki}}} \bibnamefont{and}
  \bibinfo{author}{\bibfnamefont{W.}~\bibnamefont{{Hu}}},
  \bibinfo{journal}{\prd} \textbf{\bibinfo{volume}{{\rm in press}}},
  \bibinfo{pages}{astro} (\bibinfo{year}{2007}), \eprint{astro-ph/0702278}.

\bibitem[{\citenamefont{Zhang}(2006)}]{Zha05}
\bibinfo{author}{\bibfnamefont{P.}~\bibnamefont{Zhang}},
  \bibinfo{journal}{Phys. Rev.} \textbf{\bibinfo{volume}{D73}},
  \bibinfo{pages}{123504} (\bibinfo{year}{2006}), \eprint{astro-ph/0511218}.

\bibitem[{\citenamefont{Starobinsky}(1980)}]{Sta80}
\bibinfo{author}{\bibfnamefont{A.~A.} \bibnamefont{Starobinsky}},
  \bibinfo{journal}{Phys. Lett.} \textbf{\bibinfo{volume}{B91}},
  \bibinfo{pages}{99} (\bibinfo{year}{1980}).

\bibitem[{\citenamefont{Faulkner
  et~al.}(2006{\natexlab{b}})\citenamefont{Faulkner, Tegmark, Bunn, and
  Mao}}]{FauTegBun06}
\bibinfo{author}{\bibfnamefont{T.}~\bibnamefont{Faulkner}},
  \bibinfo{author}{\bibfnamefont{M.}~\bibnamefont{Tegmark}},
  \bibinfo{author}{\bibfnamefont{E.~F.} \bibnamefont{Bunn}}, \bibnamefont{and}
  \bibinfo{author}{\bibfnamefont{Y.}~\bibnamefont{Mao}}
  (\bibinfo{year}{2006}{\natexlab{b}}), \eprint{astro-ph/0612569}.

\bibitem[{\citenamefont{Vikman}(2005)}]{Vik04}
\bibinfo{author}{\bibfnamefont{A.}~\bibnamefont{Vikman}},
  \bibinfo{journal}{Phys. Rev.} \textbf{\bibinfo{volume}{D71}},
  \bibinfo{pages}{023515} (\bibinfo{year}{2005}), \eprint{astro-ph/0407107}.

\bibitem[{\citenamefont{{Hu}}(2005)}]{Hu04c}
\bibinfo{author}{\bibfnamefont{W.}~\bibnamefont{{Hu}}}, \bibinfo{journal}{\prd}
  \textbf{\bibinfo{volume}{71}}, \bibinfo{pages}{047301}
  (\bibinfo{year}{2005}), \eprint{astro-ph/0410680}.

\bibitem[{\citenamefont{Guo et~al.}(2005)\citenamefont{Guo, Piao, Zhang, and
  Zhang}}]{Guoetal04}
\bibinfo{author}{\bibfnamefont{Z.-K.} \bibnamefont{Guo}},
  \bibinfo{author}{\bibfnamefont{Y.-S.} \bibnamefont{Piao}},
  \bibinfo{author}{\bibfnamefont{X.-M.} \bibnamefont{Zhang}}, \bibnamefont{and}
  \bibinfo{author}{\bibfnamefont{Y.-Z.} \bibnamefont{Zhang}},
  \bibinfo{journal}{Phys. Lett.} \textbf{\bibinfo{volume}{B608}},
  \bibinfo{pages}{177} (\bibinfo{year}{2005}), \eprint{astro-ph/0410654}.

\bibitem[{\citenamefont{Amendola and Tsujikawa}(2007)}]{Amendola:2007nt}
\bibinfo{author}{\bibfnamefont{L.}~\bibnamefont{Amendola}} \bibnamefont{and}
  \bibinfo{author}{\bibfnamefont{S.}~\bibnamefont{Tsujikawa}}
  (\bibinfo{year}{2007}), \eprint{arXiv:0705.0396 [astro-ph]}.

\bibitem[{\citenamefont{Dolgov and Kawasaki}(2003)}]{DolKaw03}
\bibinfo{author}{\bibfnamefont{A.~D.} \bibnamefont{Dolgov}} \bibnamefont{and}
  \bibinfo{author}{\bibfnamefont{M.}~\bibnamefont{Kawasaki}},
  \bibinfo{journal}{Phys. Lett.} \textbf{\bibinfo{volume}{B573}},
  \bibinfo{pages}{1} (\bibinfo{year}{2003}), \eprint{astro-ph/0307285}.

\bibitem[{\citenamefont{Seifert}(2007)}]{Sei07}
\bibinfo{author}{\bibfnamefont{M.~D.} \bibnamefont{Seifert}}
  (\bibinfo{year}{2007}), \eprint{gr-qc/0703060}.

\bibitem[{\citenamefont{{Bahcall} et~al.}(2005)\citenamefont{{Bahcall},
  {Serenelli}, and {Basu}}}]{BahSerBas05}
\bibinfo{author}{\bibfnamefont{J.~N.} \bibnamefont{{Bahcall}}},
  \bibinfo{author}{\bibfnamefont{A.~M.} \bibnamefont{{Serenelli}}},
  \bibnamefont{and} \bibinfo{author}{\bibfnamefont{S.}~\bibnamefont{{Basu}}},
  \bibinfo{journal}{\apjl} \textbf{\bibinfo{volume}{621}}, \bibinfo{pages}{L85}
  (\bibinfo{year}{2005}), \eprint{arXiv:astro-ph/0412440}.

\bibitem[{\citenamefont{{Vernazza} et~al.}(1981)\citenamefont{{Vernazza},
  {Avrett}, and {Loeser}}}]{VerAvrLoe81}
\bibinfo{author}{\bibfnamefont{J.~E.} \bibnamefont{{Vernazza}}},
  \bibinfo{author}{\bibfnamefont{E.~H.} \bibnamefont{{Avrett}}},
  \bibnamefont{and} \bibinfo{author}{\bibfnamefont{R.}~\bibnamefont{{Loeser}}},
  \bibinfo{journal}{\apjs} \textbf{\bibinfo{volume}{45}}, \bibinfo{pages}{635}
  (\bibinfo{year}{1981}).

\bibitem[{\citenamefont{{Sittler} and {Guhathakurta}}(1999)}]{SitGuh99}
\bibinfo{author}{\bibfnamefont{E.~C.} \bibnamefont{{Sittler}},
  \bibfnamefont{Jr.}} \bibnamefont{and}
  \bibinfo{author}{\bibfnamefont{M.}~\bibnamefont{{Guhathakurta}}},
  \bibinfo{journal}{\apj} \textbf{\bibinfo{volume}{523}}, \bibinfo{pages}{812}
  (\bibinfo{year}{1999}).

\bibitem[{\citenamefont{Will}(2006)}]{Will:2005va}
\bibinfo{author}{\bibfnamefont{C.~M.} \bibnamefont{Will}},
  \bibinfo{journal}{Living Reviews in Relativity} \textbf{\bibinfo{volume}{9}}
  (\bibinfo{year}{2006}),
  \urlprefix\url{http://www.livingreviews.org/lrr-2006-3}.

\bibitem[{\citenamefont{Klypin et~al.}(2002)\citenamefont{Klypin, Zhao, and
  Somerville}}]{KlyZhaSom02}
\bibinfo{author}{\bibfnamefont{A.}~\bibnamefont{Klypin}},
  \bibinfo{author}{\bibfnamefont{H.}~\bibnamefont{Zhao}}, \bibnamefont{and}
  \bibinfo{author}{\bibfnamefont{R.~S.} \bibnamefont{Somerville}},
  \bibinfo{journal}{Astrophys. J.} \textbf{\bibinfo{volume}{573}},
  \bibinfo{pages}{597} (\bibinfo{year}{2002}), \eprint{astro-ph/0110390}.

\end{thebibliography}

\end{document}